\documentclass[prl,aps,amsmath,amssymb,amsfonts,twocolumn,nofootinbib,longbibliography,superscriptaddress]{revtex4-2}
\usepackage{amssymb}
\usepackage{bm,mathrsfs}
\usepackage{graphicx}
\usepackage{epsfig}
\usepackage{amsmath,bbm}
\usepackage{amsfonts,amssymb}
\usepackage{times}
\usepackage{verbatim}
\usepackage[sort&compress]{natbib}
\usepackage{amsmath}
\usepackage{bm}
\usepackage{float}
\usepackage{booktabs}
\usepackage{colortbl}
\usepackage{graphicx}
\usepackage{amsmath}
\usepackage{amssymb}
\usepackage{array}
\usepackage{makecell}
\usepackage{multirow}

\allowdisplaybreaks[4]
\usepackage[colorlinks,breaklinks,linkcolor=blue,anchorcolor=blue,citecolor=blue,urlcolor=blue]{hyperref}

\usepackage{booktabs}
\definecolor{lightgray}{gray}{0.9}
\begin{document}
	\title{Variational Quantum Algorithm for Unitary Dilation}
\author{S. X. Li}
 \affiliation{Center for Quantum Science and School of Physics, Northeast Normal University, Changchun 130024,  China}

\author{Keren Li}
\affiliation{College of Physics and Optoelectronic Engineering, Shenzhen University, Shenzhen 518060, China}
\affiliation{Quantum Science Center of Guangdong-Hong Kong-Macao Greater Bay Area (Guangdong), Shenzhen 518045, China}

\author{J. B. You}
\affiliation{Quantum Innovation Centre (Q.InC), Agency for Science, Technology and Research (A*STAR), 2 Fusionopolis Way, \#08-03 Innovis, Singapore 138634}
\affiliation{Institute of High Performance Computing (IHPC), Agency for Science, Technology and Research (A*STAR), 1 Fusionopolis Way, \#16-16 Connexis, Singapore 138632}

\author{Y.-H. Chen}
\email{yehong.chen@fzu.edu.cn}
\affiliation{Fujian Key Laboratory of Quantum Information and Quantum Optics, Fuzhou University, Fuzhou 350116, China}
\affiliation{Department of Physics, Fuzhou University, Fuzhou 350116, China}
\affiliation{Quantum Information Physics Theory Research Team, Center for Quantum Computing, RIKEN, Wako-shi, Saitama 351-0198, Japan}

\author{Clemens Gneiting}	
\affiliation{Quantum Information Physics Theory Research Team, Center for Quantum Computing, RIKEN, Wako-shi, Saitama 351-0198, Japan}

\author{Franco Nori}	
\affiliation{Quantum Information Physics Theory Research Team, Center for Quantum Computing, RIKEN, Wako-shi, Saitama 351-0198, Japan}
\affiliation{Department of Physics, University of Michigan, Ann Arbor, Michigan 48109-1040, USA}

\author{X. Q. Shao}
\email{shaoxq644@nenu.edu.cn}
\affiliation{Center for Quantum Science and School of Physics, Northeast Normal University, Changchun 130024, China}
\affiliation{Institute of Quantum Science and Technology, Yanbian University, Yanji 133002, China}
    
\date{\today}
\begin{abstract}
We introduce a hybrid quantum-classical framework for efficiently implementing approximate unitary dilations of non-unitary operators with enhanced noise resilience. The method embeds a target non-unitary operator into a subblock of a unitary matrix generated by a parameterized quantum circuit with universal expressivity, while a classical optimizer adjusts circuit parameters under the global unitary constraint. As a representative application, we consider the non-unitary propagator of a Lindbladian superoperator acting on the vectorized density matrix, which is relevant for simulating open quantum systems. We further validate the approach experimentally on superconducting devices in the Quafu quantum cloud computing cluster. Compared with standard dilation protocols, our method significantly reduces quantum resource requirements and improves robustness against device noise, achieving high-fidelity simulation. Its generality also enables compatibility with non-Markovian dynamics and Kraus-operator-based evolutions, providing a practical pathway for the noise-resilient simulation of non-unitary processes on near-term quantum hardware.
\end{abstract}

\maketitle

{\it Introduction}---Quantum computers inherently operate through unitary transformations, which makes them powerful tools for simulating closed quantum systems. However, many processes of fundamental and practical importance, such as open quantum systems, dissipative dynamics, and measurement-induced evolution, are governed by non-unitary dynamics. This discrepancy raises a central question: how can unitary-based quantum computers faithfully simulate non-unitary physics?

A rigorous framework to address this problem is provided by unitary dilation, which embeds a non-unitary operator (with singular values bounded by unity) into a larger unitary matrix \cite{Levy2014Dilation,Dey2022Generalized}. The essential idea is that instead of implementing the non-unitary operator directly, one realizes its dynamics as a subblock of a physically executable unitary. This guarantees universal access to non-unitary processes on quantum hardware. While conceptually powerful, conventional dilation methods generally require large ancilla spaces or deep circuit decompositions, making them challenging to implement on near-term quantum devices~\cite{Hu2020Quantum,PhysRevResearch.4.023216,PhysRevA.106.022424,PhysRevA.106.022414,PhysRevLett.131.120401,doi:10.1021/acs.jctc.3c00316,10821275,doi:10.1137/22M1484298,https://doi.org/10.1111/sapm.70047,PhysRevA.110.012445}.
The challenge becomes particularly pressing in the noisy intermediate-scale quantum (NISQ) era \cite{doi:10.1126/science.abe8770,chow2021ibm,madsen2022quantum,wurtz2023aquila,PhysRevLett.134.090601,abughanem2025ibm}. Current platforms, typically consisting of tens of qubits with modest gate fidelities, have demonstrated capabilities in quantum simulation and computational tasks~\cite{doi:10.1126/science.abo6587,graham2022multi,PhysRevLett.131.150601,doi:10.1021/acs.jctc.3c00319,kim2023evidence,zheng2025quantum}.
Yet, without error correction, noise accumulation in deep circuits imposes strict limits on realizable algorithms. This motivates the development of hardware-efficient and noise-resilient strategies for implementing unitary dilation.

In this work, we propose a Variational Quantum Algorithm for Unitary Dilation (VQAUD). In this approach, a parameterized quantum circuit is optimized so that a designated subblock encodes the desired non-unitary operator. This hybrid quantum-classical framework directly connects the mathematical construction of dilations with their physical implementation. Crucially, by substantially reducing the number of single- and two-qubit entangling gates compared to standard dilation schemes, VQAUD makes the simulation of non-unitary processes experimentally feasible on noisy intermediate-scale devices. Beyond the computational efficiency, our approach provides a practical route to study the physics of open quantum systems, dissipative state engineering, and non-Hermitian dynamics under realistic hardware constraints. This establishes VQAUD as a versatile tool for probing non-unitary quantum phenomena with near-term quantum computers.

\begin{figure}
\includegraphics[width=1\linewidth]{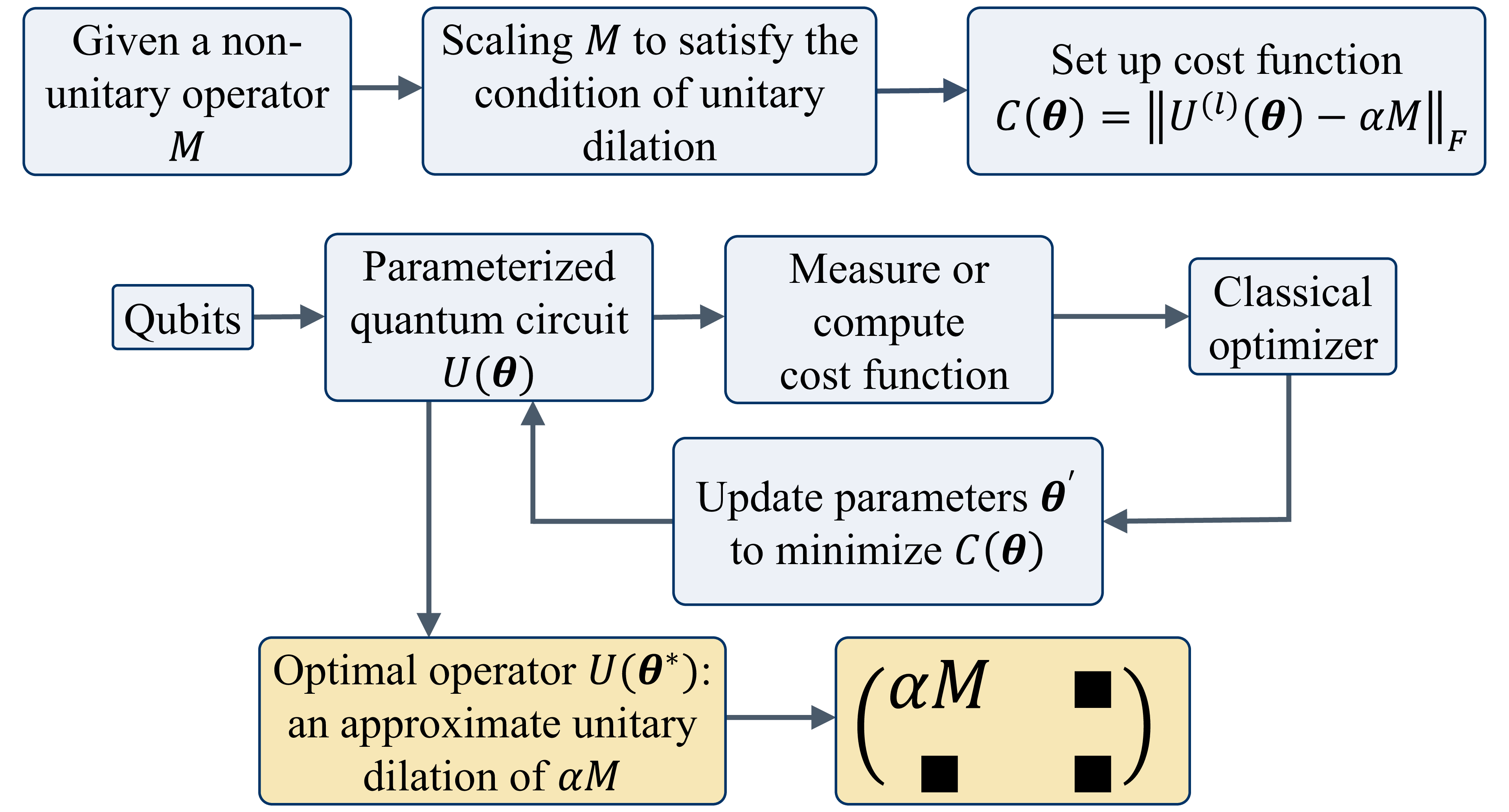}
\caption{The VQAUD allows the parameterized quantum circuit to achieve universal matrix expressivity for any given dimension. By leveraging quantum operator properties while preserving global unitary constraints, we define the cost function $C(\boldsymbol{\theta}) = \| U^{(l)}(\boldsymbol{\theta}) - \alpha M  \|_F$ and employ classical optimization algorithms to minimize it, such that the upper-left submatrix of the quantum circuit's unitary matrix approximates the target non-unitary operator $\alpha M$. This formally establishes a unitary dilation for the non-unitary operator $\alpha M$ under the given dimensional constraints.}\label{fig1}
\end{figure}

{\it VQAUD framework}---Our VQAUD approach can be divided into two parts: sustained unitarity guarantee and subblock approximation, with the main workflow illustrated in Fig.~\ref{fig1}. In general, VQAs represent a class of hybrid quantum-classical computational frameworks that combine parameterized quantum circuits with classical optimization routines~\cite{LIU2024116494,PRXQuantum.4.010328,PhysRevA.104.052409}. The core architecture comprises three fundamental elements: (I) a parameterized quantum circuit $U(\boldsymbol{\theta})$ with universal expressivity based on adjustable parameters $\boldsymbol{\theta} = (\theta_1, \ldots, \theta_p)$, (II) a cost function quantifying computational objectives, and (III) a classical optimization algorithm for parameter tuning.
The quantum circuit $U(\boldsymbol{\theta})$ consists of quantum gates with parameter-dependent operations, typically expressed as
$
U(\boldsymbol{\theta}) = \prod_{k=1}^{L} \exp(-i \theta_k H_k),
$
where $H_k$ are Hermitian generators and $\theta_k$ represent adjustable rotation angles. The circuit structure may utilize different ansatz designs depending on the target application. In subsequent iterations of the classical optimizer, the operator implemented by the quantum circuit will always maintain unitarity, regardless of parameter variations. This enables its subblocks to variationally approximate the effective non-unitary evolution of open quantum systems, thereby bridging the gap between mathematical dilation and physical implementation.

To implement the non-unitary operator $M \in \mathbb{C}^{l \times l}$ in quantum hardware, we employ the following subblock approximation strategy. First, to satisfy the mathematical condition for unitary dilation, we multiply $M$ by a scaling factor $\alpha \in (0, 1]$, ensuring that the maximum singular value of $\alpha M$ does not exceed 1. It is necessary to check the singular value decomposition of
$
\alpha M = U \Sigma V^\dagger,
$
where $(U, V)\in\mathbb{C}^{l \times l}$ are unitary matrices and $\Sigma = \mathrm{diag}(\sigma_1, \dots, \sigma_l)$ contain the singular values $\sigma_i \leq 1$.  
The central objective is to approximate the rescaled operator $\alpha M$ embedded in the subblock of a dilated unitary operator with minimal dimension $2l$. To quantify the approximation accuracy, we introduce the Frobenius-norm-based cost function:
\[
C(\boldsymbol{\theta}) = \left\| U^{(l)}(\boldsymbol{\theta}) - \alpha M \right\|_F,
\]
where $U^{(l)}(\boldsymbol{\theta})$ denotes the $l \times l$ top-left subblock of $U(\boldsymbol{\theta})$, and the Frobenius norm $\|X\|_F$ is defined as $\|X\|_F = \left(\sum_{i,j} |X_{ij}|^2\right)^{1/2}$.  
In the Supplementary Material (SM) \cite{SM}, we also considered the use of a cost function computable by quantum devices through SWAP-test-based measurements \cite{Flammia2011Direct,Wiebe2014Hamiltonian,Buhrman2001Quantum} of quantum states. Minimizing this cost function ensures that the approximated operator faithfully reproduces the target operator's dynamics while preserving the unitary structure of the entire circuit.

Next, a classical optimizer is employed to find the optimal parameters that minimize the cost function. In this work, the Broyden-Fletcher-Goldfarb-Shanno (BFGS) algorithm \cite{nocedal1999numerical}, a gradient-based method known for its efficiency in high-dimensional parameter spaces, is employed. The BFGS algorithm iteratively updates the parameter set $\boldsymbol{\theta}$ by approximating the inverse Hessian matrix $\mathcal{H}^{-1}(\boldsymbol{\theta})$ using only gradient information, thus avoiding the computational expense of calculating exact second derivatives. At each optimization step $k$, the parameter update follows
$
\boldsymbol{\theta}_{k+1} = \boldsymbol{\theta}_k - \eta_k \mathcal{H}_k^{-1} \nabla C(\boldsymbol{\theta}_k),
$
where $\nabla C(\boldsymbol{\theta}_k)$ denotes the gradient of the cost function with respect to the circuit parameters at step $k$, and $\eta_k$ is a step size determined through line search. The inverse Hessian approximation $\mathcal{H}_k^{-1}$ is recursively updated by
\[
\mathcal{H}_{k+1}^{-1} = \left(I - \frac{\boldsymbol{s}_k \boldsymbol{y}_k^T}{\boldsymbol{y}_k^T \boldsymbol{s}_k}\right)\mathcal{H}_k^{-1}\left(I - \frac{\boldsymbol{y}_k \boldsymbol{s}_k^T}{\boldsymbol{y}_k^T \boldsymbol{s}_k}\right) + \frac{\boldsymbol{s}_k \boldsymbol{s}_k^T}{\boldsymbol{y}_k^T \boldsymbol{s}_k},
\]
with $\boldsymbol{s}_k = \boldsymbol{\theta}_{k+1} - \boldsymbol{\theta}_k$ representing the parameter displacement and $\boldsymbol{y}_k = \nabla C(\boldsymbol{\theta}_{k+1}) - \nabla C(\boldsymbol{\theta}_k)$ the gradient difference. The algorithm converges to optimized parameters $\boldsymbol{\theta}^*$ that minimize the Frobenius norm difference between the circuit's subblock $U^{(l)}(\boldsymbol{\theta}^*)$ and the target operator $\alpha M$, terminating when the optimization progress becomes negligible according to predefined convergence thresholds. The resulting optimal unitary operator $U(\boldsymbol{\theta}^*)$ thus provides both the desired high-quality operator approximation in its subblock and guaranteed persistent unitarity preservation, achieving the approximate unitary dilation of the non-unitary operator $\alpha M$.

\begin{figure}
\includegraphics[width=1\linewidth]{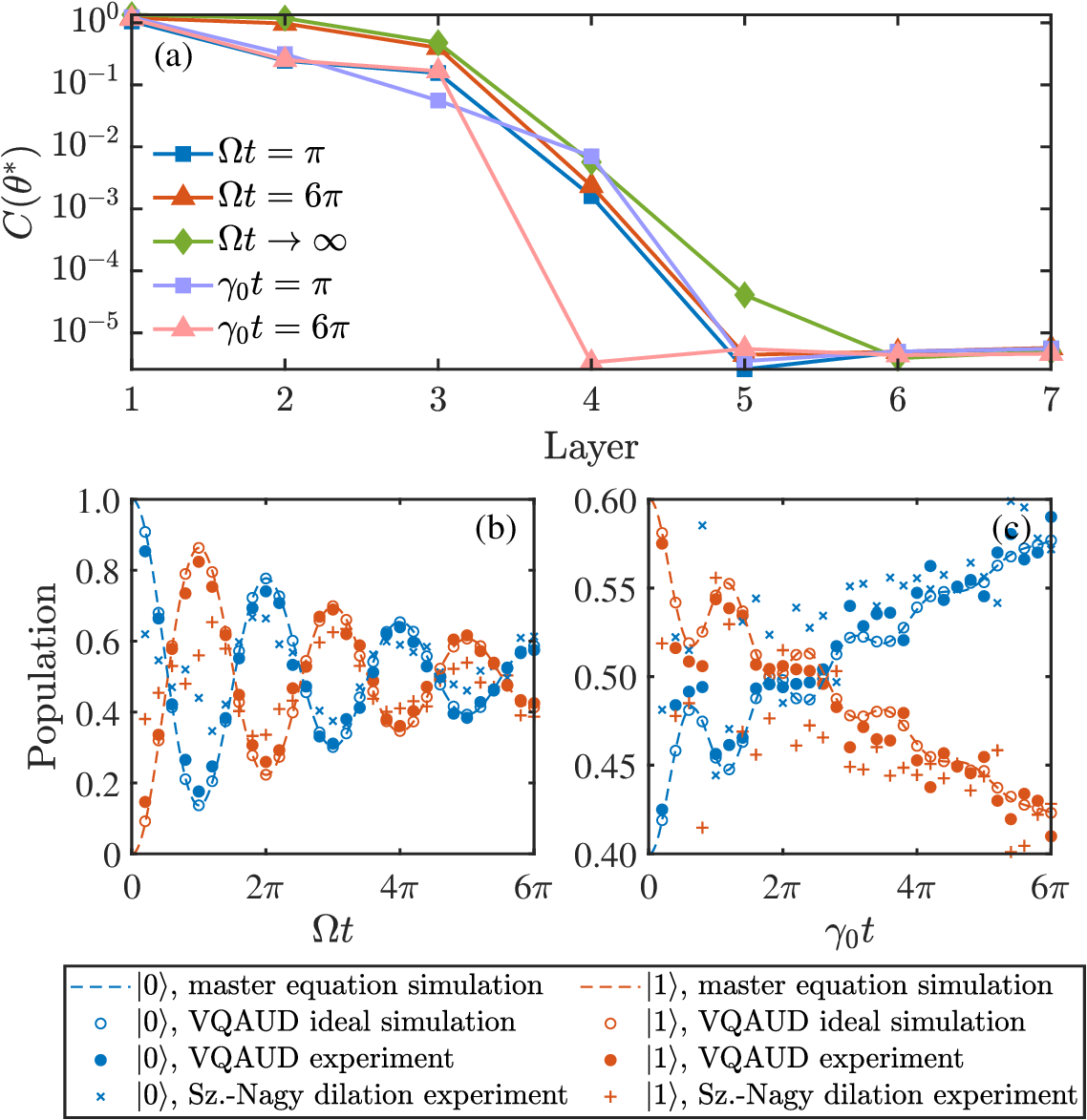}
\caption{VQAUD-based simulations of Lindblad dynamics for multiple two-level systems.
    (a) Cost functions obtained from BFGS optimization  at different circuit depths, demonstrating the implementation accuracy of the non-unitary propagator. 
    (b) Population dynamics of a Markovian driven two-level open quantum system with \( H = \frac{\Omega}{2} (|0\rangle\langle 1| + |1\rangle\langle 0|) \), decay rate \( \gamma = \Omega/10 \), and pure dephasing rate \( \gamma_{\mathrm{dp}} = \Omega/50 \), starting from the initial state \( |0\rangle \). 
    (c) Population dynamics of a non-Markovian detuned damped Jaynes--Cummings model with \( \lambda = \gamma_0/5 \) and \( \Delta = 8\lambda \), initialized in \( \rho(0) = (2|0\rangle\langle 0| + 3|1\rangle\langle 1|)/5 \). The dashed lines and hollow circles represent numerical simulation results from the master equation and VQAUD, respectively, while solid circles with plus and cross markers denote results from VQAUD and the standard Sz.-Nagy dilation algorithm implemented on the Baihua superconducting quantum computing platform. Each time-step point is estimated from \( 2^{15} \) projective measurement shots.
    }\label{fig2}
\end{figure}

{\it Simulation of open quantum systems}---One important application of our scheme lies in simulating open quantum systems. In recent years, unitary-circuit-based algorithms for open-system dynamics, which explicitly account for system-environment interactions, have been extensively investigated \cite{PhysRevA.106.022424,10.1063/5.0242648,PhysRevResearch.6.023263,https://doi.org/10.1002/qute.202400240,Suri2023twounitary,PhysRevA.109.052224,PhysRevLett.131.120401,PhysRevResearch.7.023076,PRXQuantum.3.040308,D4CP01669F,doi:10.1021/acsomega.3c09720,Hu2022generalquantum,chen2022global,Hu2020Quantum,PhysRevResearch.4.023216,PhysRevA.106.022414,PhysRevResearch.6.013135,doi:10.1021/acs.jctc.3c00316,doi:10.1021/acs.chemrev.4c00428,PhysRevLett.127.270503,PRXQuantum.3.010320,PhysRevD.106.054508,doi:10.1021/acs.jpclett.4c00576,Mahdian_2020,PhysRevResearch.3.013182,PhysRevLett.125.010501,PhysRevResearch.2.043289,PhysRevA.111.052612,kato2025exponentiallyaccurateopenquantum,2cx4-b82c}. These approaches employ a variety of mathematical frameworks for characterizing open quantum dynamics, including Lindblad master equations, Kraus operators, superoperators, and stochastic Schr\"odinger equations. Each provides a distinct representation of the system-environment interaction. Our scheme is applicable to frameworks that can be expressed as operators acting on vectors, such as the Kraus operator method and the Lindblad superoperator formalism. In the following, we focus on the Lindblad superoperator approach as a representative example, the applicability to the Kraus method is discussed in the SM \cite{SM}.

Under the Born-Markov approximation, the dynamics of an open quantum system are governed by the Lindblad master equation, 
\[ \frac{d\rho}{dt} = -i[\hat{H},\rho] + \sum_k \gamma_k( \hat{C}_k\rho\hat{C}_k^\dagger - \frac{1}{2}\{\hat{C}_k^\dagger\hat{C}_k, \rho\} ),\]
where $\rho$ is the system's density matrix, $\hat{H}$ is the Hamiltonian, and $\hat{C}_k$ are the collapse operators with associated decay rates $\gamma_k$. Here, $\{\cdot,\cdot\}$ denotes the anti-commutator ($\{A,B\}=AB+BA$). 
To implement this dynamics on a quantum circuit, we adopt the vectorization formalism, mapping the density matrix $\rho$ to a column vector $|\rho\rangle\rangle \equiv \mathrm{vec}(\rho) = (\rho_{11},\rho_{21},\dots,\rho_{nn})^T$. This transforms the Lindblad equation into the superoperator form, $d|\rho\rangle\rangle/dt = \mathcal{L}|\rho\rangle\rangle$, where the Lindbladian superoperator $\mathcal{L}$ acts on the vectorized density matrix. The formal solution is $|\rho(t)\rangle\rangle = \exp(\mathcal{L}t)|\rho(0)\rangle\rangle$. Explicitly, $\mathcal{L}$ can be expressed using Kronecker products as 
\begin{align}
\mathcal{L} ={}& -i\bigl(\mathbb{I} \otimes \hat{H} - \hat{H}^T \otimes \mathbb{I}\bigr) \nonumber\\
&+ \sum_k \Bigl[
\hat{C}_k^* \otimes \hat{C}_k 
- \frac{1}{2}\bigl(
\mathbb{I} \otimes \hat{C}_k^\dagger \hat{C}_k 
+ \hat{C}_k^T \hat{C}_k^* \otimes \mathbb{I}
\bigr)
\Bigr],\nonumber
\end{align}
where $^*$, $^T$, and $^\dagger$ denote complex conjugation, transpose, and adjoint, respectively. For an $n$-dimensional system, the vectorization enlarges the Hilbert space from $n$ to $n^2$, as $\mathcal{L}$ becomes an $n^2 \times n^2$ matrix.  
The non-unitary propagator $\exp(\mathcal{L}t)$ poses a fundamental challenge for quantum-circuit implementation because quantum gates are inherently unitary. To address this, we set $M = \exp(\mathcal{L}t)$ and apply the unitary dilation procedure described above. The target non-unitary dynamics is then encoded in the first block column of $U(\boldsymbol{\theta}^*)$ via $U(\boldsymbol{\theta}^*)(|\rho(0)\rangle\rangle \oplus |\mathbf{0}_d\rangle) = \alpha |\rho(t)\rangle\rangle \oplus |\star\rangle$, where $|\mathbf{0}_d\rangle \equiv (0,\dots,0)^T \in \mathbb{C}^d$, and $|\star\rangle$ represent auxiliary components in the dilated space that do not encode physical information. Any necessary rescaling or renormalization factor $\alpha$ can be applied via classical post-processing to recover the original Lindblad evolution.

The procedure is illustrated through a simple two-level system with Hamiltonian  
$
H = \frac{\Omega}{2} \left( |0\rangle\langle1| + |1\rangle\langle0| \right),
$
subject to spontaneous emission  
$
C_1 = \sqrt{\gamma}\,|0\rangle\langle1|
$  
and pure dephasing  
$
C_2 = \sqrt{\gamma_{\rm dp}}\left(|1\rangle\langle1| - |0\rangle\langle0|\right).
$ 
Using the parameterized circuit ansatz presented in the SM~\cite{SM}, VQAUD approximates the rescaled dynamical map \(\alpha M\). Figure~\ref{fig2}(a) displays the approximation error as a function of evolution time and circuit depth. As expected, increasing the depth improves the fidelity of the approximation. In this parameter regime, a four-layer circuit already achieves near-optimal performance. Figure~\ref{fig2}(b) compares the time evolution obtained from the four-layer VQAUD (hollow circles) with the exact Lindblad solution (dashed lines), showing excellent agreement. The VQAUD reproduces both the dissipative population decay and the coherent Rabi oscillations, demonstrating its ability to accurately capture open-system dynamics.

To further examine the versatility of the approach, we apply it to a more complex, non-Markovian scenario: a detuned, non-Markovian damped Jaynes–Cummings model governed by the fourth-order time-convolutionless master equation~\cite{PhysRevLett.109.170402,reddy2025characterizingnonmarkoviandynamicsopen,PhysRevA.81.042103,breuer2002theory},  
\[
\frac{d\rho}{dt} = -\frac{i}{2} S(t)\left[\sigma_+\sigma_-,\rho\right] 
+ \gamma(t)\left(\sigma_-\rho\sigma_+ - \frac{1}{2}\{\sigma_+\sigma_-,\rho\}\right),
\]
where \(S(t)\) and \(\gamma(t)\) denote the time-dependent Lamb shift and decay rate (explicit forms are given in Ref.~\cite{breuer2002theory} and the SM). In this model, \(\gamma(t)\) can temporarily take negative values, signaling information backflow and leading to non-Markovian revivals. Figure~\ref{fig2}(c) shows that the four-layer VQAUD (hollow circles) accurately captures these revivals and matches the theoretical predictions (dashed lines) across the entire time range considered.

Experimental validation was carried out on a superconducting quantum processor (Baihua)~\cite{Quafu2025}, where the optimal unitary-dilated propagator was implemented at selected times. For comparison, we also realized the standard Sz.-Nagy dilation by decomposing  
\[
U_{\text{dil}} = 
\begin{pmatrix}
\alpha M & -U\sqrt{\mathbb{I}-\Sigma^2}\,U^\dagger\\[4pt]
V\sqrt{\mathbb{I}-\Sigma^2}\,V^\dagger & V\Sigma U^\dagger
\end{pmatrix}
\]
into quantum gates (explicit decompositions are provided in the SM~\cite{SM}). Experimental data (solid circles) show small deviations from ideal simulations (dashed lines) due to hardware noise. Nevertheless, VQAUD consistently outperforms the Sz.-Nagy implementation (plus and cross markers), primarily owing to its reduced circuit complexity. Additional calibration data and implementation details can be found in the SM~\cite{SM}.

\begin{table*}
  \centering
  \caption{Quantum resource comparison for implementing $M = \exp({\mathcal{L}t})$ in Figs~\ref{fig2}--\ref{fig4}. 
  VQAUD {\it significantly reduces gate counts} compared to previous methods. 
  Reduction percentages are shown relative to Ref.~\cite{PhysRevResearch.4.023216} and Ref.~\cite{doi:10.1021/acs.jctc.3c00316}.}
  \label{tab:cnot_counts}
  \renewcommand\arraystretch{1.3}
  \begin{tabular}{
    c|
    >{\columncolor{lightgray}}c c c >{\columncolor{lightgray}}c|
    >{\columncolor{lightgray}}c c c >{\columncolor{lightgray}}c|
    >{\columncolor{lightgray}}c c c  
    }
    \toprule
    \multirow{2}{*}{\makecell{\textbf{Simulated} \\ \textbf{systems}}}
    & \multicolumn{4}{c|}{\textbf{Single-qubit gate count}}
    & \multicolumn{4}{c|}{\textbf{Two-qubit gate count}}
    & \multicolumn{3}{c}{\textbf{Qubit count}} \\
    \cmidrule(lr){2-5} \cmidrule(lr){6-9} \cmidrule(lr){10-12}
    & \textbf{VQAUD} & Ref.~\cite{PhysRevResearch.4.023216} & Ref.~\cite{doi:10.1021/acs.jctc.3c00316} & \makecell{\textbf{Reduction} (\%$\downarrow$)}
    & \textbf{VQAUD} & Ref.~\cite{PhysRevResearch.4.023216} & Ref.~\cite{doi:10.1021/acs.jctc.3c00316} & \makecell{\textbf{Reduction} (\%$\downarrow$)}
    & \textbf{VQAUD} & Ref.~\cite{PhysRevResearch.4.023216} & Ref.~\cite{doi:10.1021/acs.jctc.3c00316} \\
    \midrule
    2-level 
    & \textbf{36} & 287 & 74 & (87.5\% / 51.4\%)
    & \textbf{8}  & 84  & 20 & (90.5\% / 60.0\%)
    & \textbf{3}  & 4 & 3 \\
    3-level 
    & \textbf{110} & 5490 & 1378 & (98.0\% / 92.0\%)
    & \textbf{25}  & 1789 & 444  & (98.6\% / 94.4\%)
    & \textbf{5}   & 6 & 5 \\
    4-level 
    & \textbf{206} & 5479 & 1377 & (96.2\% / 85.0\%)
    & \textbf{49}  & 1779 & 444  & (97.2\% / 89.0\%)
    & \textbf{5}   & 6 & 5 \\
    \bottomrule
  \end{tabular}
\end{table*}

\begin{figure}
\includegraphics[width=1\linewidth]{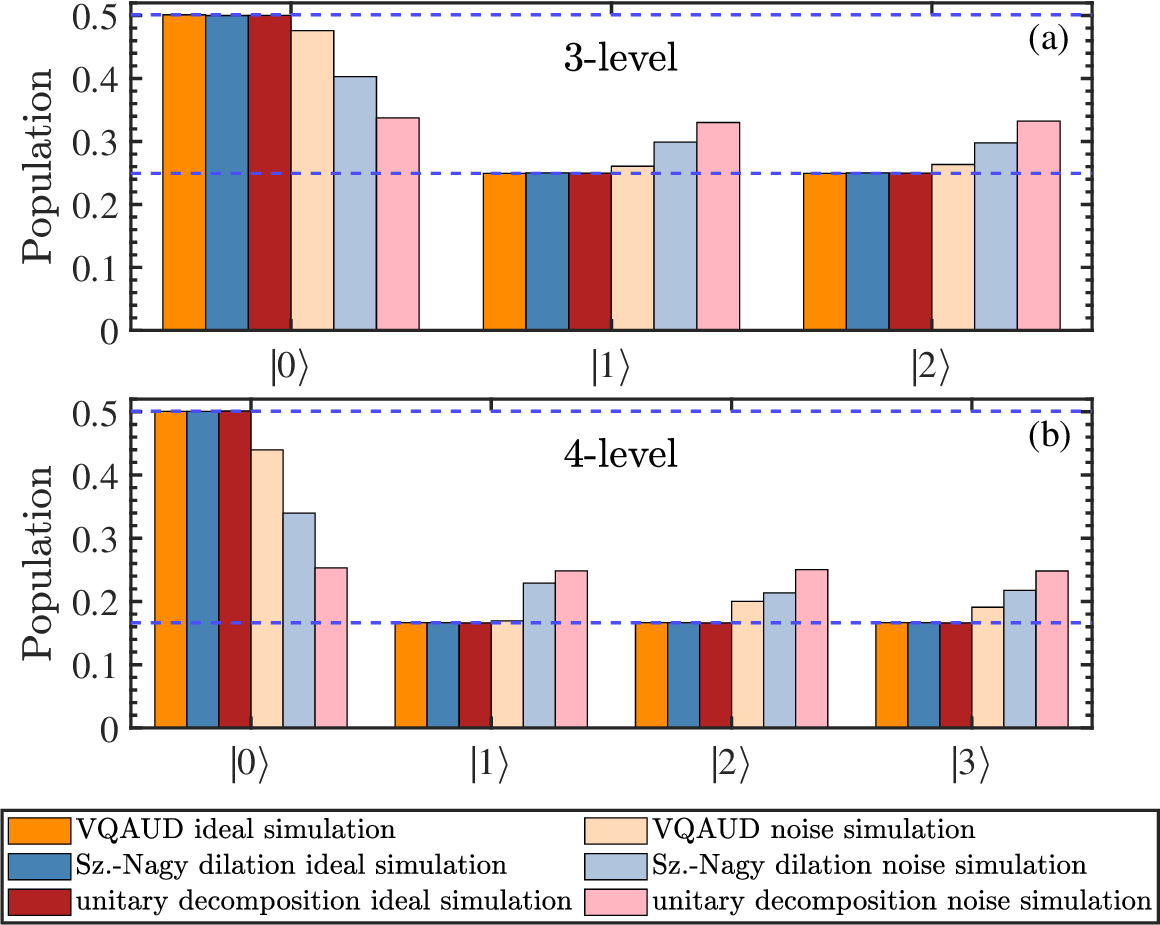}
\caption{Numerical simulation results of steady states ($ t\to\infty$) of (a) three-level and (b) four-level open systems ($\gamma=\Omega/10$) from initial states in uniform superposition via three approaches (VQAUD, standard Sz.-
Nagy dilation algorithm and unitary decomposition algorithm) using quantum devices noise parameters $\lambda=2\times10^{-3}$ and $\omega=1\times10^{-3}$. The dashed lines indicate the ideal reference from the master equation simulation.
}\label{fig4}
\end{figure}

{\it Extension to multilevel open quantum systems---}We next investigate the performance of VQAUD in more complex settings involving three- and four-level open quantum systems. 
In current superconducting platforms, qubit connectivity is typically limited, making it challenging to implement fully connected entangling circuits required by theoretical dilation schemes. 
To address this constraint and emulate realistic hardware conditions, here we adopt a numerical approach that incorporates composite noise channels into all logical gates, effectively simulating quantum circuit execution on noisy processors.  Specifically, for each qubit \(j\) participating in a logical gate, we apply a phase-damping channel
$
\mathcal{N}_{\text{dep}}^{(j)}[\rho] = (1-\lambda)\rho + \lambda Z_j \rho Z_j
$
and an amplitude-damping channel
$
\mathcal{N}_{\text{amp}}^{(j)}[\rho] = K_0(\omega)\,\rho\,K_0^\dagger(\omega) + K_1(\omega)\,\rho\,K_1^\dagger(\omega),
$
with Kraus operators
\[
K_0(\omega) = 
\begin{pmatrix} 
1 & 0 \\ 
0 & \sqrt{1-\omega} 
\end{pmatrix}, \qquad  
K_1(\omega) = 
\begin{pmatrix} 
0 & \sqrt{\omega} \\ 
0 & 0 
\end{pmatrix}.
\]
The parameters \(\lambda\) and \(\omega\) quantify the strengths of dephasing and amplitude-damping noise, respectively, and characterize the detrimental effects of gate errors. 

VQAUD for these systems is implemented using a five-qubit parameterized circuit.
We adopt a single-layer, parallel architecture in which each single-qubit gate carries two rotation parameters and is followed by enriched non-nearest-neighbor entangling operations (see SM~\cite{SM}).
This ansatz offers a favorable trade-off between expressivity and trainability, enhancing the circuit’s representational power while mitigating the onset of barren plateaus~\cite{larocca2025barren}.

For benchmarking, we compare VQAUD with two alternative schemes. In addition to the standard Sz.-Nagy dilation, we consider a unitary-decomposition approach~\cite{PhysRevResearch.4.023216}, which separates the target operator \(M\) into its Hermitian and anti-Hermitian components, \(M=\hat{S}+\hat{A}\). The operator can then be expressed as a linear combination of unitaries,
\[
M = \lim_{\epsilon \to 0} \frac{1}{2\epsilon} 
\left( i e^{-i\epsilon \hat{S}} - i e^{i\epsilon \hat{S}} + e^{\epsilon \hat{A}} - e^{-\epsilon \hat{A}} \right),
\]
allowing direct implementation on quantum processors, since each exponential term corresponds to a unitary operation.

Building on the above noise models, we simulate the steady-state dynamics by evaluating \(M = \lim_{t \to \infty} \exp(\mathcal{L}t)\) in a V-type three-level system and in a four-level configuration where a single ground state couples to three excited states. Figure~\ref{fig4} summarizes the results obtained using the three methods. Under ideal, noiseless conditions, all methods yield comparable performance. However, when realistic gate errors are included, both the standard Sz.-Nagy and unitary-decomposition approaches exhibit significant performance degradation, whereas VQAUD maintains high accuracy. This demonstrates the superior robustness of VQAUD against gate imperfections, underscoring its practical utility for noisy quantum hardware.

Table~\ref{tab:cnot_counts} presents a comparison of the quantum resources required to implement \(M = \exp(\mathcal{L} t)\) for two- to four-level systems. VQAUD achieves dramatic reductions in both single- and two-qubit gate counts relative to standard dilation and unitary-decomposition methods. The savings become more pronounced as system dimensions increase, while the total qubit overhead remains minimal and comparable to reference methods. These results demonstrate that VQAUD provides a highly efficient and scalable approach for simulating open-system dynamics, reducing circuit depth and entangling operations without sacrificing fidelity.

\begin{figure}
\includegraphics[width=1\linewidth]{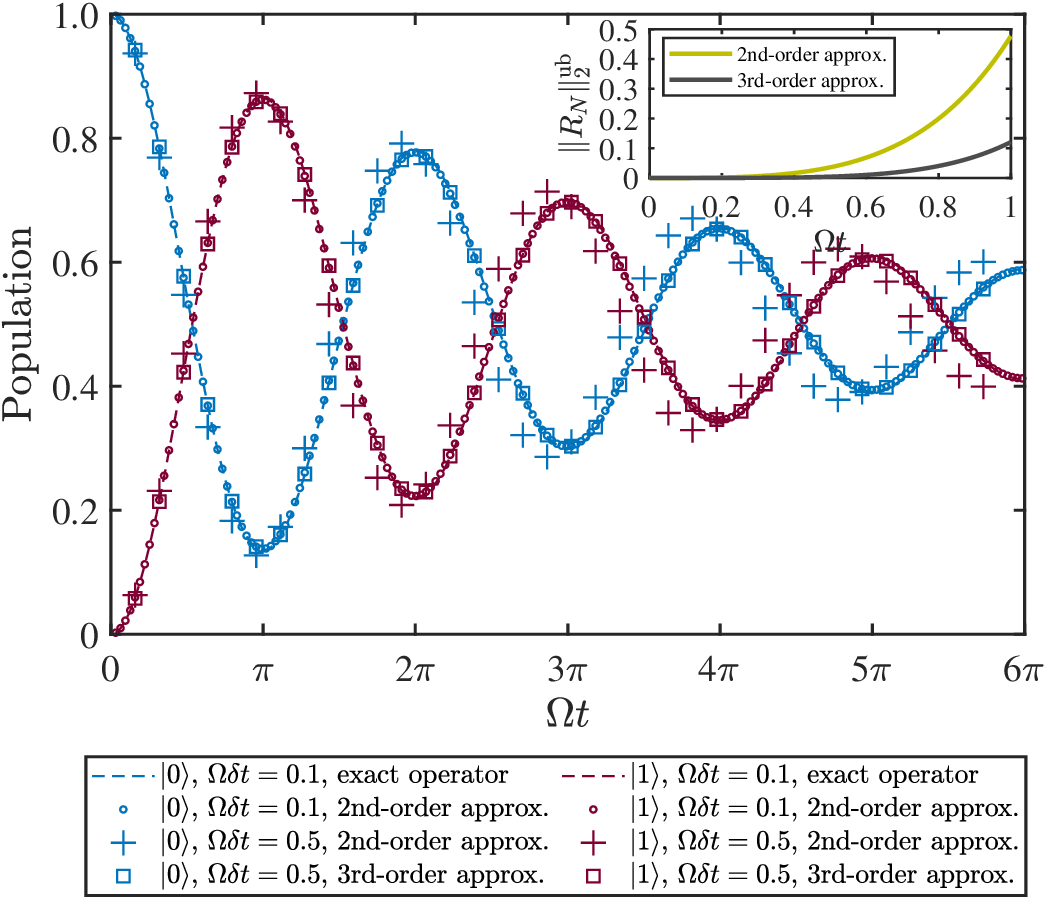}
\caption{Single-optimization VQAUD simulation of time-independent Lindblad dynamics for a driven two-level Markovian open quantum system. The inset shows the upper bound of the 2-norm of the remainder in the Taylor expansion of the Liouvillian propagator at different approximation orders.
}\label{fig_taylor}
\end{figure}

{\it Single optimization for time-independent dynamics}---For time-independent master equations, the VQAUD protocol is optimized for a small initial time step $\delta t$, after which the dynamics at later times are efficiently obtained by repeatedly applying the same optimized circuit for integer multiples of $\delta t$. The corresponding quantum circuits are shown in the SM~\cite{SM}. This single-optimization strategy avoids re-optimizing at each time step and already yields accurate results, as demonstrated by the dashed lines in Fig.~\ref{fig_taylor}, where the evolution of the previously discussed two-level system matches the master equation simulation in Fig.~\ref{fig2}(b). Furthermore, the Liouvillian propagator can be approximated using a short-time Taylor expansion $\exp(\mathcal{L} \delta t)= \sum_{n=0}^{N} (\mathcal{L} \delta t)^n/n! + R_N(\delta t)$, where the operator 2-norm of the remainder satisfies $\|R_N(\delta t)\|_2 \le [\|\mathcal{L}  \delta t\|_2^{N+1}/(N+1)!] \exp({\|\mathcal{L} \delta t\|_2})$. The inset of Fig.~\ref{fig_taylor} shows how this upper bound depends on $\Omega \delta t$ for second- and third-order expansions. For $\Omega t = 0.1$, the second-order, single-optimization VQAUD results (hollow circles) closely follow the exact evolution, confirming both the validity of the approximation and the efficiency of the protocol. At larger time steps (e.g., $\Omega t = 0.5$), the second-order expansion fails (plus markers), whereas higher-order expansions (square markers) remain accurate. Importantly, in higher-dimensional systems where classical computation of the Liouvillian propagator is intractable, this approach enables quantum computers to directly simulate the master equation without explicitly constructing the full propagator or relying on the auxiliary quantum devices described in the SM \cite{SM}, thereby extending its applicability to large-scale open quantum systems.

{\it Conclusions}---We have proposed a noise-robust VQAUD that implements non-unitary operators via optimized quantum circuit synthesis. By recasting the non-unitary propagator of Lindbladian dynamics as a rescaled subblock of a dilated unitary, our framework enables accurate and efficient simulation of open quantum systems. The hybrid quantum-classical architecture uses variational optimization to synthesize compact unitary representations that approximate target non-unitary maps while substantially reducing quantum-resource overhead.

We demonstrate VQAUD across two- to four-level open quantum systems and identify two principal advantages. First, VQAUD markedly reduces single- and two-qubit gate counts relative to standard dilation protocols---an effect that becomes more pronounced for higher-dimensional systems. Second, the variationally optimized circuits exhibit enhanced robustness to realistic NISQ noise, improving fidelity when gate errors and decoherence are present. Moreover, VQAUD is broadly extensible: it naturally accommodates non-Markovian dynamics and Kraus-operator simulations and can be generalized to many-body platforms such as spin chains and cavity arrays.

The choice of parameterized circuit (ansatz) is critical for practical performance. Different ans\"atze may offer comparable expressivity yet differ substantially in noise resilience, so careful circuit design is essential. Promising future directions include automated circuit-architecture search \cite{Du2022Quantum} to optimize ans\"atze without increasing depth, and the incorporation of Schr\"odinger-Heisenberg quantum circuits \cite{PhysRevLett.131.060406} that interleave real and virtual layers within a hybrid operational scheme. We expect that, as hardware advances toward higher-fidelity multi-qubit gates, improved connectivity, and more effective error-mitigation/correction, VQAUD will provide a practical pathway for realizing high-accuracy simulations of non-unitary processes on near- and mid-term quantum devices.

{\it Note added}---We recently became aware of Ref.~\cite{yyln-q22s}, which develops a stochastic Magnus expansion method for robust variational quantum simulation of Lindblad dynamics. Their approach reformulates non-unitary evolution into a stochastic framework, whereas our method relies on a direct unitary dilation to encode the Lindblad dynamics. These two approaches are complementary and highlight the growing interest in variational strategies for simulating open quantum systems.

{\it Acknowledgments}---This work was supported by the National Natural Science
Foundation of China (NSFC) under Grant No. 12174048.
Y.-H.C. was supported by the National Natural Science Foundation of China under Grant No. 12304390, the Fujian 100 Talents Program, and the Fujian Minjiang Scholar Program.
F.N. is supported in part by: 
the Japan Science and Technology Agency (JST) 
[via the CREST Quantum Frontiers program Grant No. JPMJCR24I2, 
the Quantum Leap Flagship Program (Q-LEAP), and the Moonshot R\&D Grant Number JPMJMS2061].

\textit{Data availability}---The data that support the findings of this article are openly
 available~\cite{li_2025_17385836}.

 \bibliography{manuscript.bbl}

\end{document}


\title{Supplemental Material:\\Variational Quantum Algorithm for Unitary Dilation}
\author{S. X. Li}
 \affiliation{Center for Quantum Science and School of Physics, Northeast Normal University, Changchun 130024,  China}

\author{Keren Li}
\affiliation{College of Physics and Optoelectronic Engineering, Shenzhen University, Shenzhen 518060, China}
\affiliation{Quantum Science Center of Guangdong-Hong Kong-Macao Greater Bay Area (Guangdong), Shenzhen 518045, China}

\author{J. B. You}
\affiliation{Quantum Innovation Centre (Q.InC), Agency for Science, Technology and Research (A*STAR), 2 Fusionopolis Way, \#08-03 Innovis, Singapore 138634}
\affiliation{Institute of High Performance Computing (IHPC), Agency for Science, Technology and Research (A*STAR), 1 Fusionopolis Way, \#16-16 Connexis, Singapore 138632}

\author{Y.-H. Chen}
\email{yehong.chen@fzu.edu.cn}
\affiliation{Fujian Key Laboratory of Quantum Information and Quantum Optics, Fuzhou University, Fuzhou 350116, China}
\affiliation{Department of Physics, Fuzhou University, Fuzhou 350116, China}
\affiliation{Quantum Information Physics Theory Research Team, Center for Quantum Computing, RIKEN, Wako-shi, Saitama 351-0198, Japan}

\author{Clemens Gneiting}	
\affiliation{Quantum Information Physics Theory Research Team, Center for Quantum Computing, RIKEN, Wako-shi, Saitama 351-0198, Japan}

\author{Franco Nori}	
\affiliation{Quantum Information Physics Theory Research Team, Center for Quantum Computing, RIKEN, Wako-shi, Saitama 351-0198, Japan}
\affiliation{Department of Physics, University of Michigan, Ann Arbor, Michigan 48109-1040, USA}

\author{X. Q. Shao}
\email{shaoxq644@nenu.edu.cn}
\affiliation{Center for Quantum Science and School of Physics, Northeast Normal University, Changchun 130024, China}
\affiliation{Institute of Quantum Science and Technology, Yanbian University, Yanji 133002, China}
    
\date{\today}

\maketitle

\section{Quantum state measurement-based cost function}

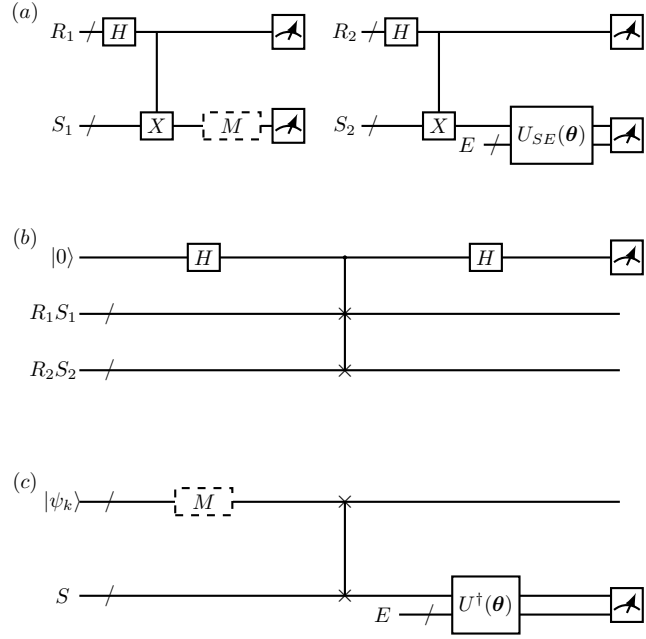
\begin{figure}
   \centerline{
   \begin{tikzpicture}[thick]
   \ctikzset{scale=2.5}
   \tikzstyle{every node}=[font=\normalsize,scale=0.8]
     \tikzstyle{operator} = [draw,shape=rectangle
     ,fill=white,minimum width=0.8em, minimum height=0.7em] 
     \tikzstyle{operator2} = [draw,shape=rectangle,fill=white,minimum width=3em, minimum height=3.em] 
     \tikzstyle{operator22} = [draw,shape=rectangle,fill=white,minimum width=3em, minimum height=9.5em] 
     \tikzstyle{operator3} = [draw,dashed, shape=rectangle,fill=white,minimum width=3em, minimum height=1em] 
     \tikzstyle{operator4} = [draw,shape=rectangle,dashed, minimum width=1.5cm, minimum height=1cm] 
     \tikzstyle{operator5} = [draw,shape=rectangle,dashed, minimum width=5.75cm, minimum height=3cm] 
     \tikzstyle{operator6} = [draw=pink,shape=rectangle,dashed, minimum width=5cm, minimum height=4cm] 
     \tikzstyle{phase} = [fill,shape=circle,minimum size=3pt,inner sep=0pt]
     \tikzstyle{surround} = [fill=blue!10,thick,draw=black,rounded corners=2mm]
     \tikzstyle{ellipsis} = [fill,shape=circle,minimum size=2pt,inner sep=0pt]
     \tikzstyle{ellipsis1} = [draw,shape=circle,minimum size=2pt,inner sep=0pt]
     \tikzstyle{ellipsis3} = [draw,shape=circle,fill=white, minimum size=2pt,inner sep=1pt]
     \tikzstyle{ellipsis4} = [draw,shape=circle,fill=white, minimum size=4pt,inner sep=3pt]
     \tikzstyle{ellipsis2} = [draw,shape=circle,minimum size=0.5pt,inner sep=0pt]
     \tikzset{xmark/.style={font=\large, inner sep=0pt},
     meter/.append style={fill=white, draw, inner sep=5, rectangle, font=\vphantom{A}, minimum width=15, 
     path picture={\draw[black] ([shift={(.05,.2)}]path picture bounding box.south west) to[bend left=40] ([shift={(-.05,.2)}]path picture bounding box.south east);\draw[black,-latex] ([shift={(0,.15)}]path picture bounding box.south) -- ([shift={(.15,-.08)}]path picture bounding box.north);}}}
     \node at (0.3,-0.9) (qin){$(a)$};
     \node at (0.5,-1.) (qin){$R_1$};
     \node at (0.55,-1.) (q3) {};
     \node[] (end3) at (1.7,-1.) {} edge [-] (q3);
     \node at (0.65,-1.) (qin){$/$};
     \node at (0.5,-1.5) (qin){$S_1$};
     \node at (0.55,-1.5) (q3) {};
     \node[] (end3) at (1.7,-1.5) {} edge [-] (q3);
     \node at (0.65,-1.5) (qin){$/$};
     \node[operator] (op22) at (0.8,-1.) {$H$};
     \coordinate (xc) at (1,-1);
     \coordinate (xd) at (1,-1.5);
     \draw[thick] (xc) -- (xd);   
     \node[operator] (Xdown) at (xd) {$X$};
     \node[operator3] (op22) at (1.4,-1.5) {$M$};
     \node[meter] at (1.7,-1.5) (qin){};
     \node[meter] at (1.7,-1.) (qin){};
     \node at (2,-1.) (qin){$R_2$};
     \node at (2.05,-1.) (q3) {};
     \node[] (end3) at (3.5,-1.) {} edge [-] (q3);
     \node at (2.15,-1.) (qin){$/$};
     \node at (2,-1.5) (qin){$S_2$};
     \node at (2.05,-1.5) (q3) {};
     \node[] (end3) at (3.5,-1.5) {} edge [-] (q3);
     \node[operator] (op22) at (2.3,-1.) {$H$};
     \node at (2.15,-1.5) (qin){$/$};
     \node at (2.65,-1.6) (qin){$E$};
     \node at (2.8,-1.6) (qin){$/$};
     \node at (2.7,-1.6) (q3) {};
     \node[] (end3) at (3.5,-1.6) {} edge [-] (q3);
     \coordinate (xc) at (2.5,-1);
     \coordinate (xd) at (2.5,-1.5);
     \draw[thick] (xc) -- (xd);   
     \node[operator] (Xdown) at (xd) {$X$};
     \node[operator2] (op22) at (3.1,-1.55) {$U_{SE}(\bm{\theta})$} ;
     \node[meter] at (3.5,-1.55) (qin){};
     \node[meter] at (3.5,-1.) (qin){};
     \node at (0.3,-2.1) (qin){$(b)$};
     \node at (0.5,-2.2) (qin){$|0\rangle$};
     \node at (0.55,-2.2) (q3) {};
     \node[] (end3) at (3.5,-2.2) {} edge [-] (q3);
     \node at (0.45,-2.5) (qin){$R_1S_1$};
     \node at (0.55,-2.5) (q3) {};
     \node[] (end3) at (3.5,-2.5) {} edge [-] (q3);
     \node at (0.75,-2.5) (qin){$/$};
     \node at (0.75,-2.8) (qin){$/$};
     \node at (0.45,-2.8) (qin){$R_2S_2$};
     \node at (0.55,-2.8) (q3) {};
     \node[] (end3) at (3.5,-2.8) {} edge [-] (q3);
     \coordinate (xc) at (2,-2.5);
     \node[ellipsis] (Xup) at (xc) {};
     \coordinate (xd) at (2,-2.8);
     \node[xmark] (Xup) at (xc) {$\times$};
     \draw[thick] (xc) -- (xd); 
     \node[xmark] (Xdown) at (xd) {$\times$};
     \coordinate (xc) at (2,-2.2);
     \node[ellipsis] (Xup) at (xc) {};
     \coordinate (xd) at (2,-2.5);
     \draw[thick] (xc) -- (xd); 
     \node[operator] (op22) at (1.25,-2.2) {$H$} ;
     \node[operator] (op22) at (2.75,-2.2) {$H$} ;
     \node[meter] at (3.5,-2.2) (qin){};
     \node at (0.3,-3.4) (qin){$(c)$};
     \node at (0.5,-3.5) (qin){$|\psi_k\rangle$};
     \node at (0.5,-4) (qin){$S$};
     \node at (2.2,-4.1) (qin){$E$};
     \node at (0.55,-3.5) (q3) {};
     \node[] (end3) at (3.5,-3.5) {} edge [-] (q3);
     \node at (0.75,-3.5) (qin){$/$};
     \node at (0.5,-4) (qin){};
     \node at (0.55,-4) (q3) {};
     \node[] (end3) at (3.5,-4) {} edge [-] (q3);
     \node at (0.75,-4) (qin){$/$};
     \node at (2.45,-4.1) (qin){$/$};
     \node at (2.25,-4.1) (q3) {};
     \node[] (end3) at (3.5,-4.1) {} edge [-] (q3);
     \node[operator3] (op22) at (1.25,-3.5) {$M$} ;
     \coordinate (xc) at (2,-3.5);
     \node[xmark] (Xup) at (xc) {$\times$};
     \coordinate (xd) at (2,-4);
     \node[xmark] (Xdown) at (xd) {$\times$};
     \draw[thick] (xc) -- (xd);     
     \node[operator2] (op22) at (2.75,-4.05) {$U^{\dagger}(\bm{\theta})$} ;
     \node[meter] at (3.5,-4.05) (qin){};
   \end{tikzpicture} } 
   \caption{Circuit diagrams for the hardware-native estimators of the cost functions. (a) Preparation of the variational Choi state $\Phi_{\boldsymbol\theta}$. The state is prepared by creating a maximally entangled state $|\Phi^+\rangle$ on registers $R_2S_2$, applying the parameterized unitary operator $U_{SE}(\boldsymbol\theta)$ on $S_2E$, and post-selecting the ancilla $E$ on the state $|0\rangle$. (b) SWAP test for estimating the fidelity $F(\Phi_{\boldsymbol\theta},\Phi_M)$ between the variational and target Choi states. A control qubit in the $|+\rangle$ state is used to apply controlled-SWAP gates between the registers ($R_1\!\leftrightarrow\!R_2$, $S_1\!\leftrightarrow\!S_2$); the overlap is given by measuring $\sigma_x$ on the control. (c) Circuit for measuring the cross-term $\langle\psi_k| \tilde M^\dagger \alpha M|\psi_k\rangle$ for the regression cost function $\mathcal L_{\mathrm{reg}}$. An ancilla qubit is used to control the application of the target operation $\alpha M$ and the variational block $\tilde M^\dagger$ on the probe state $|\psi_k\rangle$; the desired matrix element is obtained by measuring $\sigma_x$ on the ancilla.}
\label{SWAP test} 
\end{figure}

In the main text, we used the non-unitary propagator $M=\exp({\mathcal{L}t})$ of the open quantum system under investigation as an example to discuss the unitary dilation process of $M$ and its applications. However, for larger-scale problems, storing the information of the matrix $M$ classically becomes costly, since $M \in \mathbb{C}^{d \times d}$ with $d = 2^n$, meaning its dimension grows exponentially with the number of qubits. This makes it difficult to perform estimation using the original cost function. To amortize this burden and enhance scalability, we propose three potential strategies, which leverage quantum state measurement-based cost functions and the device-in-the-loop method to avoid classical simulation of quantum circuits, as illustrated in Fig.~\ref{SWAP test}. 

We reformulate the goal as follows. Given a non-unitary operator $M$, we seek a unitary operator $U_{SE}(\boldsymbol\theta)\in\mathbb{U}(2^{n+1})$ acting on an $n$-qubit system and ancilla qubits, such that the upper-left block approximates $\alpha M$:
\begin{equation}
\label{eq:block}
\tilde M(\boldsymbol\theta)
\;:=\;
\,\big\langle 0_{ a}\big|\,U_{SE}(\boldsymbol\theta)\,\big|0_{ a}\big\rangle
\;\approx\; \alpha M.
\end{equation}
This constitutes a block-encoding of \(M\) that can be executed on near-term hardware via shallow, native-gate parameterized circuits and single-qubit measurements on the ancilla. Throughout, all measurements are single-copy Pauli measurements unless stated otherwise.


The first potential strategy uses fidelity of the corresponding Choi states as the cost function, 
\begin{equation}
\mathcal{L}_{\mathrm{Choi}}= 1- F(\Phi_{\boldsymbol\theta},\Phi_M).
\end{equation}
Here, let $\Phi_M$ be the target Choi state induced by the non-unitary operator $M$ and $\Phi_{\boldsymbol\theta}$ the Choi state  induced by the variational block $\tilde M(\boldsymbol\theta)$, which are sketched in Fig.~\ref{SWAP test}(a). Specifically, the preparation of $\Phi_{\bm{\theta}}$ for the quantum circuit \(R_2S_2E\) proceeds as follows.  
We initialize two registers \(R_2S_2\) (each consisting of \(n\) qubits) and an ancillary register \(E\).  
On \(R_2S_2\), we prepare the maximally initial entangled state \(|\Phi^+\rangle = \frac{1}{\sqrt{d}} \sum_j |j\rangle_{R_2} |j\rangle_{S_2}\),  
apply the parameterized unitary operator \(U_{SE}(\boldsymbol\theta)\) to \(S_2E\),  
and finally post-select on the ancillary register \(E\) being in state \(|0\rangle\) to obtain \(\Phi_{\boldsymbol\theta}\).

Next, the cost function based on Choi states can be measured using direct fidelity estimation (DFE)~\cite{Flammia2011Direct}.
DFE estimates $F(\Phi_{\boldsymbol\theta},\Phi_M)$ by sampling Pauli strings $P\in\mathcal{P}_{2n}$ with importance weights, 
\[
F=\sum_{P\in\mathcal{P}_{2n}}\!\! w_P\,\mu_P, \quad 
w_P =\frac{\mathrm{Tr}(\Phi_M P)^2}{\sum_Q \mathrm{Tr}(\Phi_M Q)^2}, \quad 
\mu_P =\frac{\mathrm{Tr}(\Phi_{\boldsymbol\theta} P)}{\mathrm{Tr}(\Phi_M P)}.
\]

Thus, the strategy can formulate as follows. Using a small shot budget, prepare $\Phi_M$ repeatedly and estimate a sketch of $|\mathrm{Tr}(\Phi_M P)|$ over a random subset of Pauli strings.
Perform importance sampling with $\tilde w_P$ and use a self-normalized or control-variate estimator to correct residual bias. This approach retains the hardware simplicity, only preparations and Pauli measurements of $\Phi_M$, 
while recovering much of the variance reduction of ideal DFE. Notably, if $\Phi_M$ is low-weight in Pauli basis, the DFE protocol becomes particularly resource-efficient.

The second potential strategy relies on being able to prepare both Choi states on hardware. A SWAP-based overlap estimator provides a simple, fully coherent objective [Fig.~\ref{SWAP test}(b)]~\cite{Buhrman2001Quantum}.
Attach a control qubit in \(|+\rangle\), apply parallel controlled-SWAP gates between corresponding pairs
\((R_1\!\leftrightarrow\!R_2)\) and \((S_1\!\leftrightarrow\!S_2)\), then measure \(\sigma_x\) on the control.
The expectation gives 
\[1/2+\mathrm{Tr}(\Phi_{\boldsymbol\theta}\Phi_A)/{2}.\] 
Remarkably, Choi states are required to be normalized to unit trace before the SWAP test. We ignore an normalization factor which can be estimated in an additional experiment.

To circumvent the need for controlled gates in the SWAP test, we mention two strategies that offer shallower circuit implementations while providing unbiased estimates of the overlap $\mathrm{Tr}(\Phi_{\boldsymbol\theta}\Phi_A)$. One approach is a destructive SWAP test performed via Bell-basis measurements on each corresponding pair of qubits from the two Choi states, followed by classical parity processing. This method replaces coherent controlled-SWAP gates with local Bell measurements and classical post-processing, significantly reducing circuit depth. Another strategy uses two-copy randomized measurements: applying the same random local Clifford gate to both copies, measuring in the computational basis, and employing a standard overlap estimator for $\mathrm{Tr}(\rho\sigma)$. Both techniques yield unbiased estimates of $\mathrm{Tr}(\Phi_{\boldsymbol\theta}\Phi_A)$ while requiring only shallow circuits, making them suitable for near-term hardware with limited coherence times.


If Choi states are unavailable, we bring with the third potential strategy with $L_{\mathrm{reg}}$ as our cost function.
We minimize
\begin{equation}
\mathcal L_{\mathrm{reg}}(\boldsymbol\theta)=\frac{1}{K}\sum_{k=1}^K \big\|
\tilde M(\boldsymbol\theta)|\psi_k\rangle - \alpha M|\psi_k\rangle
\big\|_2^2,
\end{equation}
over a set of probe states $\{|\psi_1\rangle,|\psi_2\rangle,...,|\psi_K\rangle\}$, which may consist of computational basis states, product states, or states generated via Haar scrambling. Expanding the norm  leads to matrix elements of the form
$\langle\psi_j| \tilde M^\dagger\tilde M|\psi_j\rangle$, $\langle\psi_j| \alpha M^\dagger \alpha M|\psi_j\rangle$ which are either classically known or estimated once, and cross terms $\langle\psi_j| \tilde M^\dagger \alpha M|\psi_j\rangle$. The latter are accessible experimentally by quantum circuit shown in Fig.~\ref{SWAP test}(c) \cite{Wiebe2014Hamiltonian}.

We assume that all gate parameters in the parameterized quantum circuits support analytic parameter-shift derivatives on the hardware. Consequently, we can employ a hybrid approach that combines the two previously proposed schemes, to find the unitary dilation \(U_{SE}(\boldsymbol\theta)\), where parameterized quantum circuits are implemented on quantum devices.

\section{Details of quantum circuits}

\begin{figure*}[!htbp]
\includegraphics[width=1\linewidth]{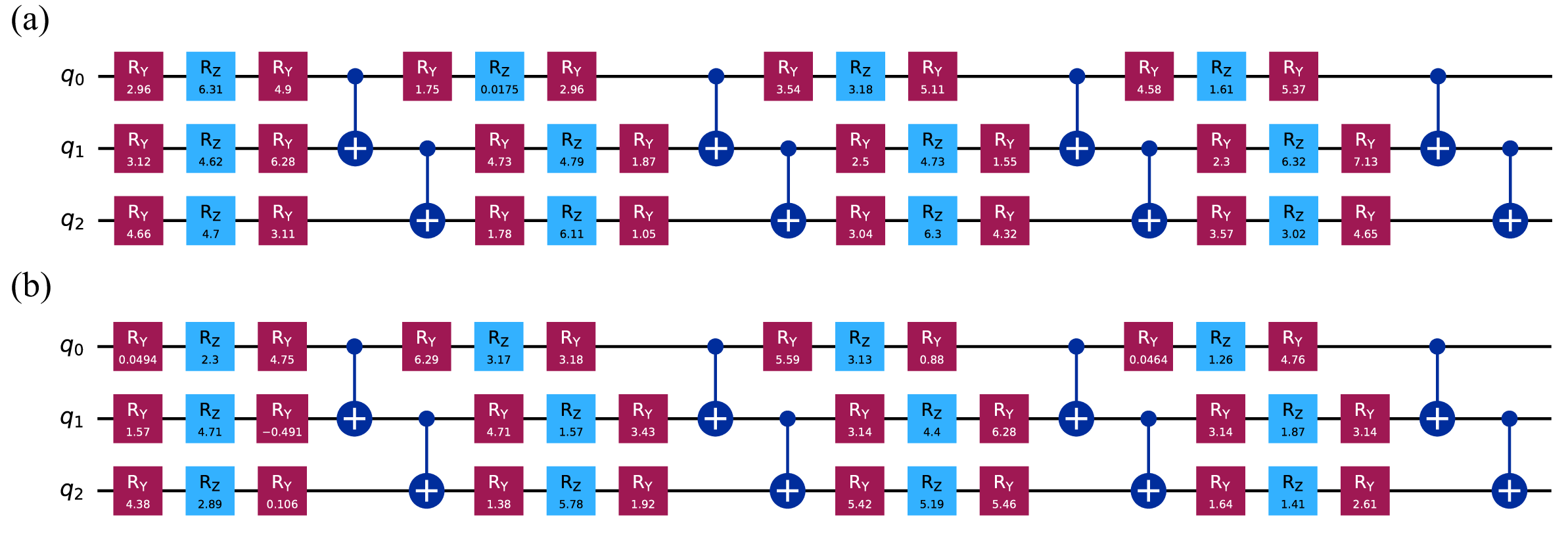}
\caption{Variational quantum circuit for simulating two-level quantum open systems. The four-layer VQA-constructed algorithm combines parameterized $R(\theta_i)$ rotations and CNOT entangling gates, optimized to approximate the (a) Markovian and (b) non-Markovian propagator with $\Omega t = 6\pi$ and $\gamma_0 t = 6\pi$ ($\alpha=0.9$), respectively. Circuit parameters are trained through the Frobenius norm minimization $\min_{\boldsymbol{\theta}}\|U^{(2)}(\boldsymbol{\theta}) - \alpha M\|_F$, achieving final approximation error (a) $2.6 \times 10^{-3}$ and (b) $3.3 \times 10^{-6}$.}\label{2_level}
\end{figure*}

\begin{figure*}[!htbp]
\includegraphics[width=1\linewidth]{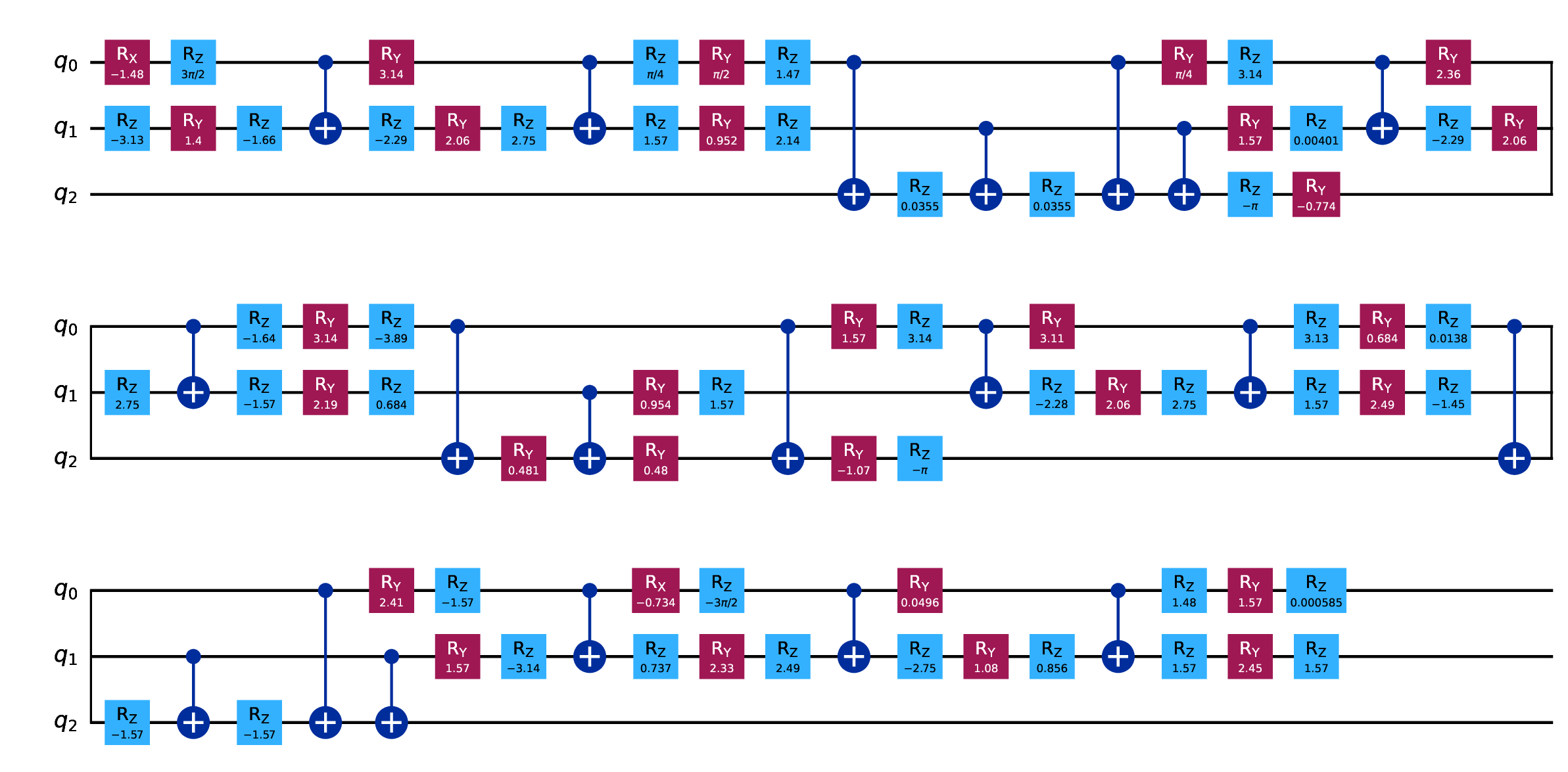}
\caption{Sz.-Nagy dilation simulation of two-level open systems in a quantum circuit. The unitary dilated propagator with $\Omega t = 6\pi$ ($\alpha=0.9$) is decomposed into a quantum circuit comprising of single-qubit gates and two-qubit gates.}\label{method2}
\end{figure*}

In this section, we present the details of the quantum circuits used in the main text. Figures~\ref{2_level} and ~\ref{method2} show the different quantum circuits containing 3 qubits to simulate two-level open quantum systems under specified time conditions using the VQA for Unitary Dilation (VQAUD) and the standard Sz.-Nagy dilation approaches, respectively. As visually evidenced, the VQAUD protocol reduces the quantum circuit complexity in unitary dilation implementation. 

In the main text, we have simulated a non-Markovian detuned damped
Jaynes--Cummings. Here, we provide the complex expressions of the Lamb shift $S(t)$ and decay rate $\gamma(t)$ used in the implementation:
\begin{eqnarray}
S(t) &=& \frac{\gamma_0 \lambda \Delta}{\lambda^2 + \Delta^2} \left[1 - e^{-\lambda t} \left(\cos(\Delta t) + \frac{\lambda}{\Delta} \sin(\Delta t)\right)\right] \nonumber\\&&
- \frac{\gamma_0^2 \lambda^2 \Delta^3 e^{-\lambda t}}{2 (\lambda^2 + \Delta^2)^3} \Bigg\{ 
    \left[1 - 3 \left(\frac{\lambda}{\Delta}\right)^2\right] \left(e^{-\lambda t} - e^{-\lambda t} \cos(2 \Delta t)\right) \nonumber
    - 2 \left[1 - \left(\frac{\lambda}{\Delta}\right)^4\right] \Delta t \sin(\Delta t) \nonumber\\&&
    + 4 \left[1 + \left(\frac{\lambda}{\Delta}\right)^2\right] \lambda t \cos(\Delta t) - \frac{\lambda}{\Delta} \left[3 - \left(\frac{\lambda}{\Delta}\right)^2\right] e^{-\lambda t} \sin(2 \Delta t) \Bigg\},
\end{eqnarray}
\begin{eqnarray}
\gamma(t) &=& \frac{\gamma_0 \lambda^2}{\lambda^2 + \Delta^2} \left[1 - e^{-\lambda t} \left(\cos(\Delta t) - \frac{\Delta}{\lambda} \sin(\Delta t)\right)\right] \nonumber\\
&& +\frac{\gamma_0^2 \lambda^5 e^{-\lambda t}}{2 (\lambda^2 + \Delta^2)^3} \Bigg\{ \left[1 - 3 \left(\frac{\Delta}{\lambda}\right)^2\right] \left(e^{-\lambda t} - e^{-\lambda t} \cos(2 \Delta t)\right) \nonumber\\
&& - 2 \left[1 - \left(\frac{\Delta}{\lambda}\right)^4\right] \lambda t \cos(\Delta t) + 4 \left[1 + \left(\frac{\Delta}{\lambda}\right)^2\right] \Delta t \sin(\Delta t) \nonumber\\
&& + \frac{\Delta}{\lambda} \left[3 - \left(\frac{\Delta}{\lambda}\right)^2\right] e^{-\lambda t} \sin(2 \Delta t) \Bigg\},
\end{eqnarray}
where $\gamma_0$, $\lambda$, and $\Delta$ represent the Markovian decay rate, the spectral width, and the detuning between the center frequency of the 
cavity and the atomic transition frequency, respectively.

\begin{figure*}[!htbp]
\includegraphics[width=1\linewidth]{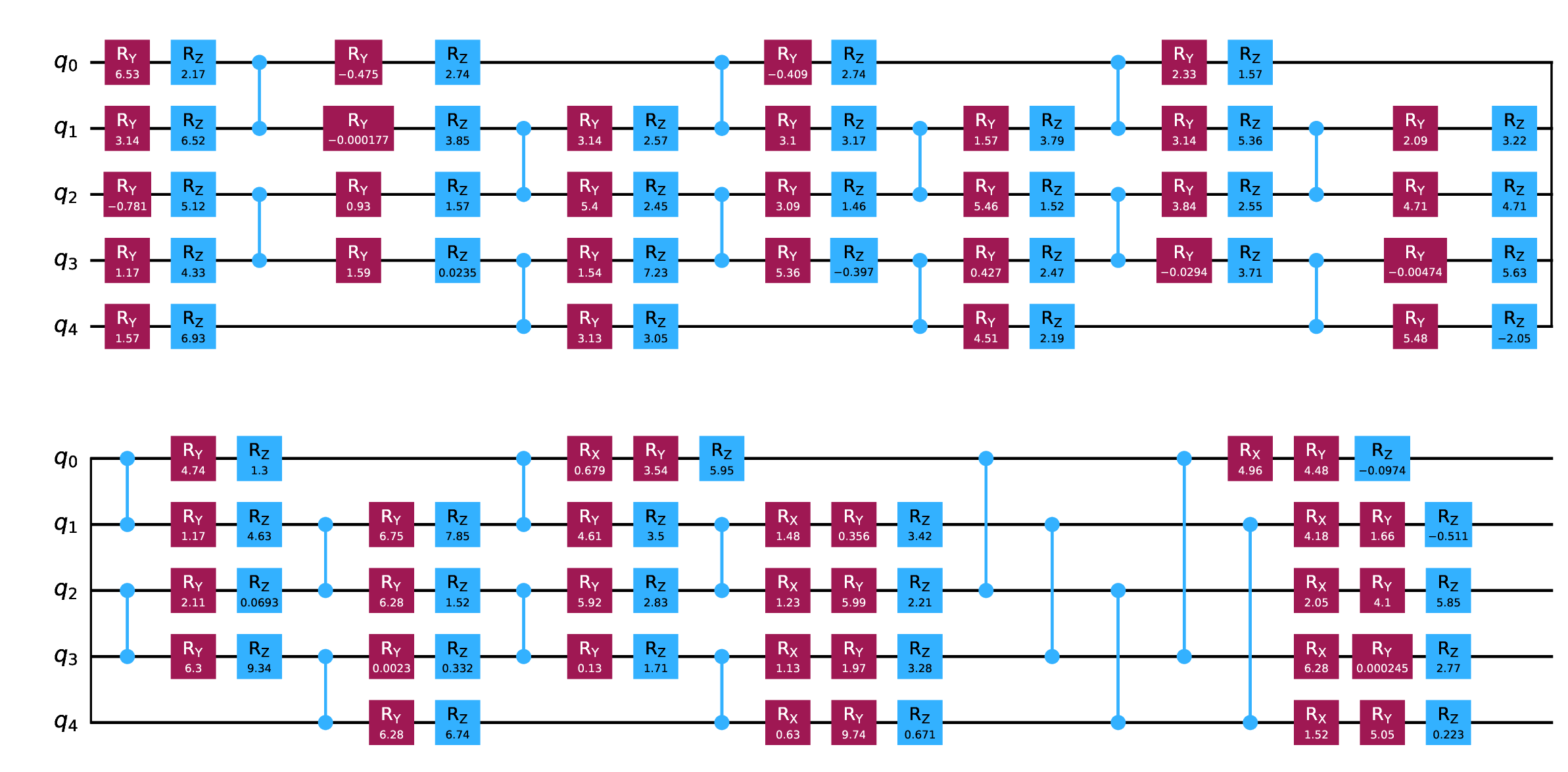}
\caption{Variational quantum circuit for steady-state simulation of three-level open systems. The VQA-constructed architecture combines parameterized $R(\theta_i)$ rotations and CZ entangling gates, optimized to approximate the propagator $ \exp({\mathcal{L}t})$ with $t \to \infty$ ($\alpha=0.5$). Circuit parameters are trained through Frobenius norm minimization $\min_{\boldsymbol{\theta}}\|U^{(9)}(\boldsymbol{\theta}) - \alpha M\|_F$,  achieving final approximation error $5.0 \times 10^{-4}$.  }\label{3_level}
\end{figure*}

\begin{figure*}[!htbp]
\includegraphics[width=1\linewidth]{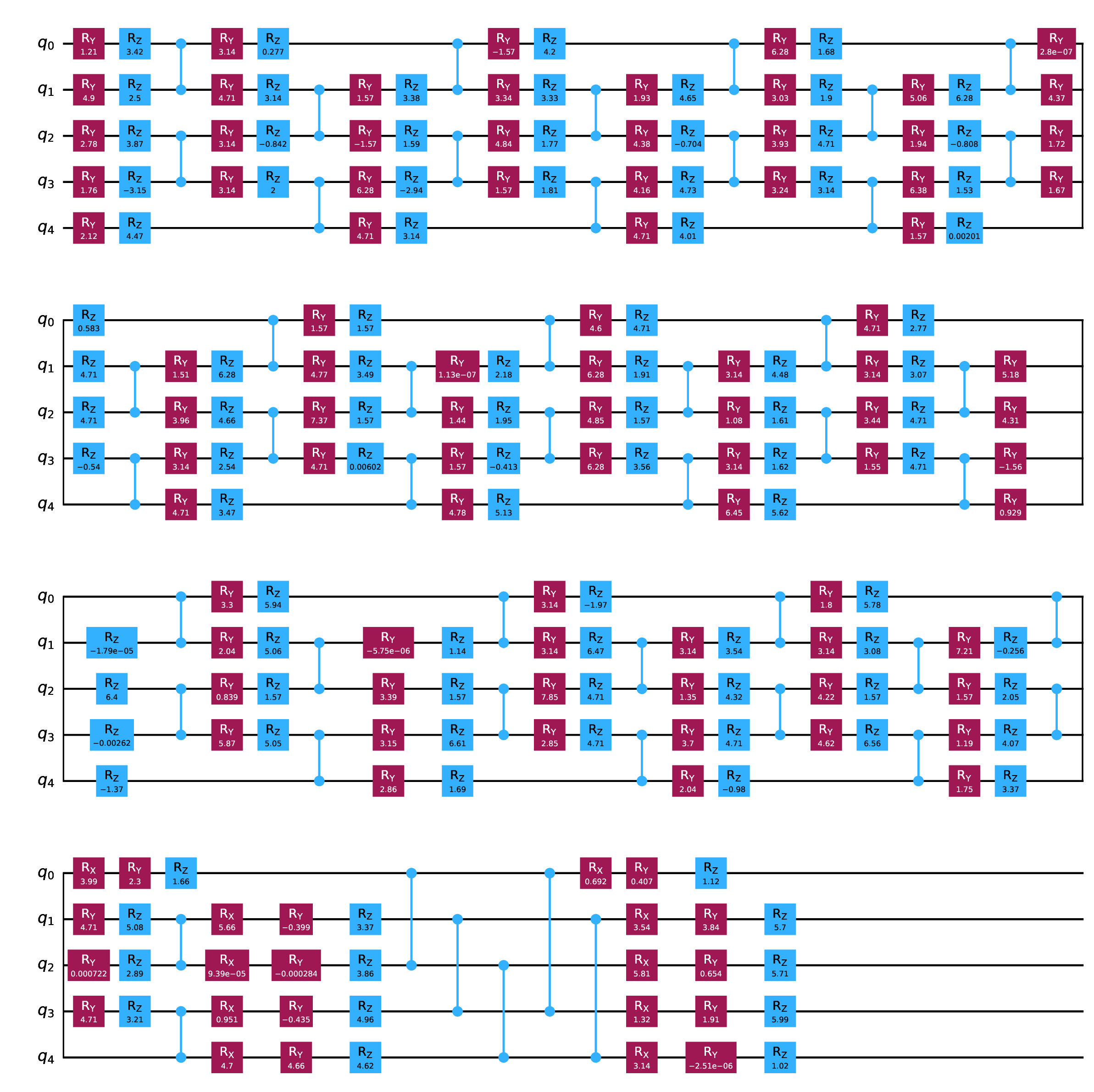}
\caption{Variational quantum circuit for steady-state simulation of four-level open systems. The VQA-constructed architecture combines parameterized $R(\theta_i)$ rotations and CZ entangling gates, optimized to approximate the propagator $\exp({\mathcal{L}t})$ with $t \to \infty$ ($\alpha=0.5$). Circuit parameters are trained through Frobenius norm minimization $\min_{\boldsymbol{\theta}}\|U^{(16)}(\boldsymbol{\theta}) - \alpha M\|_F$, achieving final approximation error $9.0 \times 10^{-5}$. }\label{4_level}
\end{figure*}

\begin{figure*}[!htbp]
\includegraphics[width=0.6\linewidth]{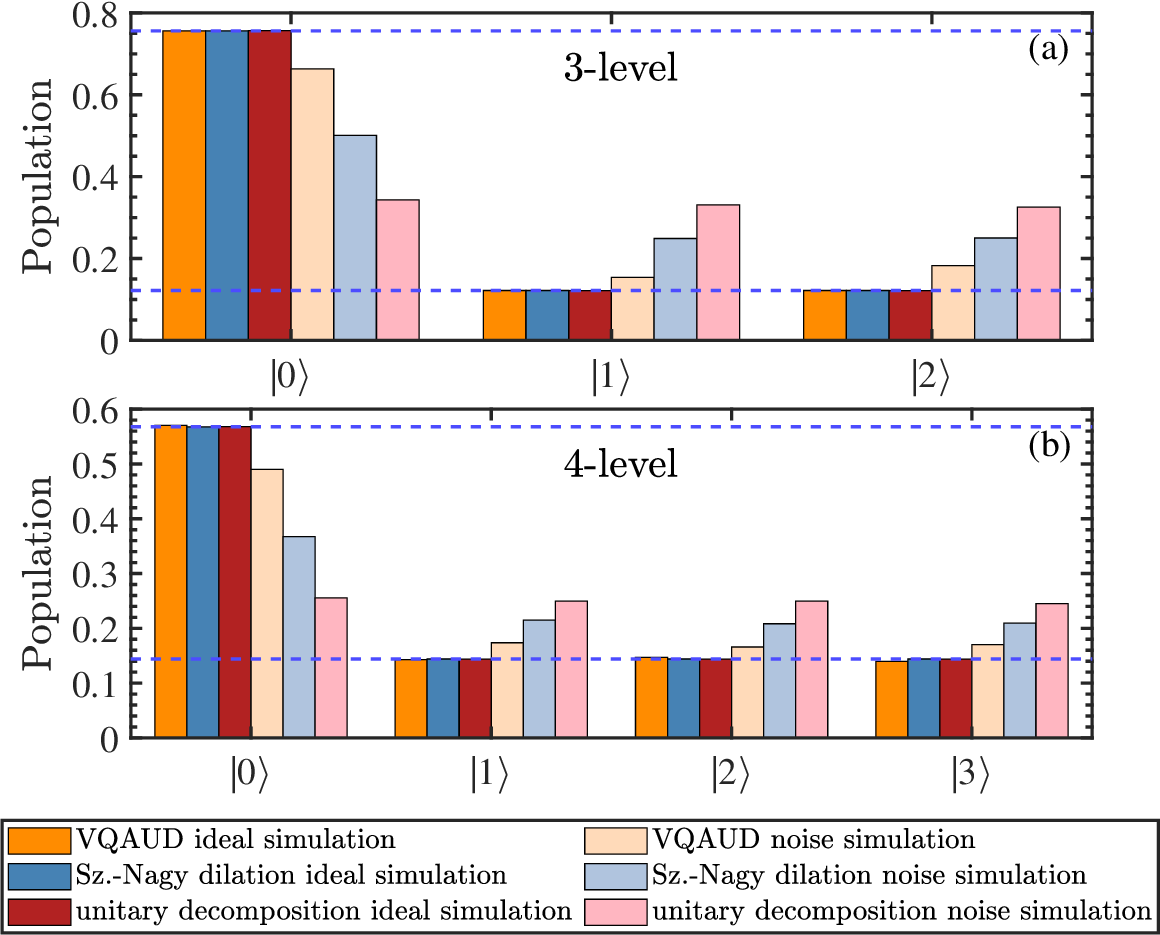}
\caption{Numerical simulation results of finite-time evolutions of (a) three-level  open system ($\Omega t=5$) from the initial state $|0\rangle$ and (b) four-level open system ($\Omega t=1$) from initial states in uniform superposition via three approaches (VQAUD, standard Sz.-
Nagy dilation algorithm, and unitary decomposition algorithm) using the quantum devices noise parameters $\lambda=1\times10^{-3}$ and $\omega=1\times10^{-3}$. The dashed lines indicate the ideal reference from the master equation simulation.}\label{finite_time}
\end{figure*}

\begin{figure*}
\includegraphics[width=1\linewidth]{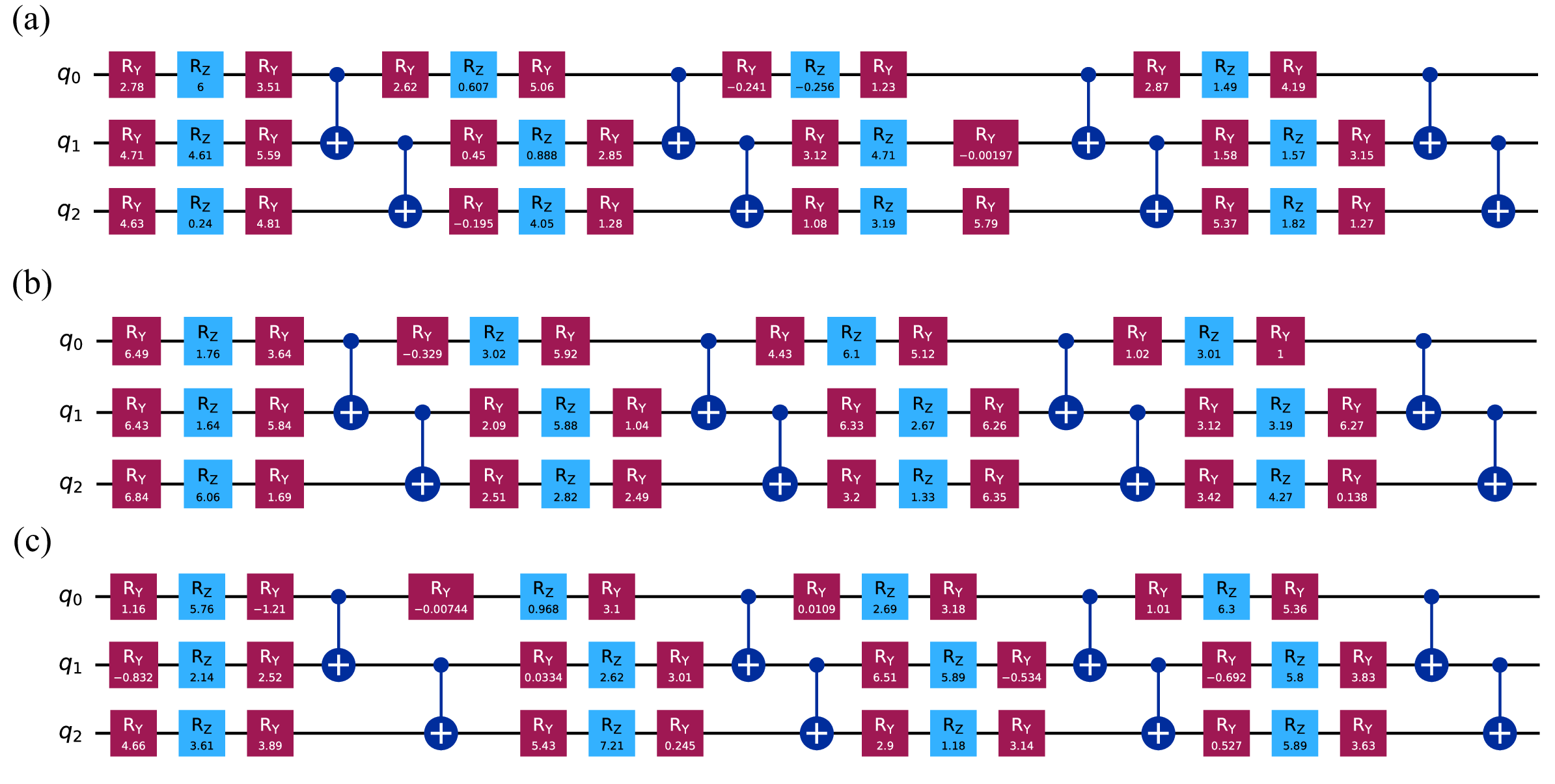}
\caption{Variational quantum circuit for simulating time-independent two-level quantum open systems within single optimization of time step (a) $\Omega t=0.1$ (2nd-order approximation), (b) $\Omega t=0.5$ (2nd-order approximation) and (c) $\Omega t=0.5$ (3rd-order approximation).}\label{taylor}
\end{figure*}

Furthermore, 5-qubit parameterized quantum circuits used to compute and implement the unitary dilation of operators at three-levels and four-levels $M= e^{\mathcal{L}t}$ $(t\to \infty)$
are presented in Figs.~\ref{3_level} and ~\ref{4_level}. Additionally, we have computed the finite-time evolution simulations of the corresponding systems, with the results shown in Fig.~\ref{finite_time}. The quantum computational resources required for these simulations are presented in Table~\ref{tab:finite time}. Figure~\ref{taylor} shows the quantum circuit for implementing single-optimization VQAUD.

\FloatBarrier

\begin{table*}[ht]
  \centering
  \caption{Quantum resource comparison for implementing $M = \exp({\mathcal{L}t})$ with finite time in Fig~\ref{finite_time}. 
  VQAUD {\it significantly reduces gate counts} compared to previous methods. 
  Reduction percentages are shown relative to Ref.~\cite{PhysRevResearch.4.023216} and Ref.~\cite{doi:10.1021/acs.jctc.3c00316}.}
  \label{tab:finite time}
  \renewcommand\arraystretch{1.3}
  \begin{tabular}{
    c|
    >{\columncolor{lightgray}}c c c >{\columncolor{lightgray}}c|
    >{\columncolor{lightgray}}c c c >{\columncolor{lightgray}}c|
    >{\columncolor{lightgray}}c c c  
    }
    \toprule
    \multirow{2}{*}{\makecell{\textbf{Simulated} \\ \textbf{systems}}}
    & \multicolumn{4}{c|}{\textbf{Single-qubit gate count}}
    & \multicolumn{4}{c|}{\textbf{Two-qubit gate count}}
    & \multicolumn{3}{c}{\textbf{Qubit count}} \\
    \cmidrule(lr){2-5} \cmidrule(lr){6-9} \cmidrule(lr){10-12}
    & \textbf{VQAUD} & Ref.~\cite{PhysRevResearch.4.023216} & Ref.~\cite{doi:10.1021/acs.jctc.3c00316} & \makecell{\textbf{Reduction} (\%$\downarrow$)}
    & \textbf{VQAUD} & Ref.~\cite{PhysRevResearch.4.023216} & Ref.~\cite{doi:10.1021/acs.jctc.3c00316} & \makecell{\textbf{Reduction} (\%$\downarrow$)}
    & \textbf{VQAUD} & Ref.~\cite{PhysRevResearch.4.023216} & Ref.~\cite{doi:10.1021/acs.jctc.3c00316} \\
    \midrule
    2-level 
    & \textbf{36} & 287 & 74 & (87.5\% / 51.4\%)
    & \textbf{8}  & 84  & 20 & (90.5\% / 60.0\%)
    & \textbf{3}  & 4 & 3 \\
    3-level 
    & \textbf{174} & 5488 & 1347 & (96.8\% / 87.1\%)
    & \textbf{41}  & 1791 & 431  & (97.7\% / 90.5\%)
    & \textbf{5}   & 6 & 5 \\
    4-level 
    & \textbf{238} & 5414 & 1381 & (95.6\% / 82.8\%)
    & \textbf{57}  & 1773 & 444  & (96.8\% / 87.2\%)
    & \textbf{5}   & 6 & 5 \\
    \bottomrule
  \end{tabular}
\end{table*}

\section{Experimental parameters of superconducting quantum computing
platform}

\begin{figure*}
\includegraphics[width=0.7\linewidth]{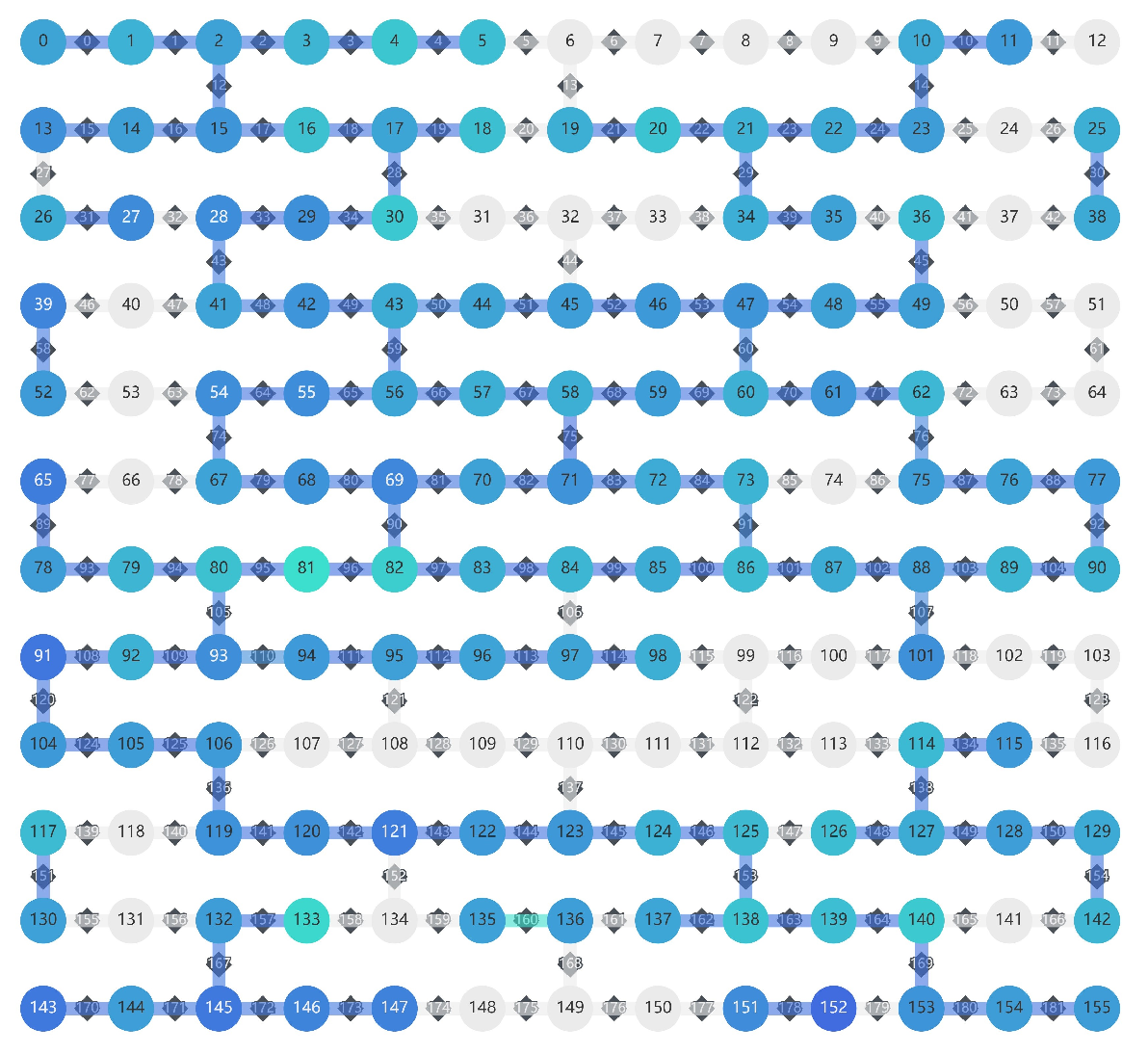}
\caption{Used superconducting quantum computing platform Baihua, where blue dots/lines denote available qubits/connections and gray dots/lines indicate unavailable qubits/connections. Baihua is developed and maintained by Superconducting Quantum Computing Group from Beijing Academy of Quantum Information Sciences.}\label{Baihua}
\end{figure*}

In this section, we provide additional details about the experimental platform used in the main text. The Baihua superconducting quantum computing platform consists of more than one hundred superconducting qubits, with the qubit information and connectivity for two-qubit entangling gates shown in Fig.~\ref{Baihua}. Table~\ref{tab:Baihua} presents the most recent calibration data for the qubits used in our experiments.

\begin{table*}
\caption{Calibration data for Baihua. The calibration date was 05/29/25.}
\label{tab:Baihua}
\centering
\footnotesize
\begin{tabular}{lccc}
\toprule
& \multicolumn{3}{c}{Qubit} \\
\cmidrule(lr){2-4}
Parameter & 21 & 22 & 23 \\
\midrule
T2 ($\mu$s)            & 40.05  & 40.27 & 50.34  \\
f (GHz)               & 3.99  & 4.43  & 4.06  \\
T1 ($\mu$s)           & 59.18  & 69.02  & 84.98 \\
Gate error ($\times10^{-4}$)  & 5.8  & 10  & 4.9  \\
Readout error ($\times10^{-2}$) & 1.85 & 2.12  & 2.73  \\
\midrule
\multicolumn{4}{l}{Multiqubit errors ($\times10^{-3}$):} \\
\cmidrule(l){1-4}
(21,22) & \multicolumn{3}{c}{7.6} \\
(22,23) & \multicolumn{3}{c}{8.9} \\
\bottomrule
\end{tabular}

\end{table*}

\section{Kraus method simulation}

\begin{figure*}
\includegraphics[width=0.6\linewidth]{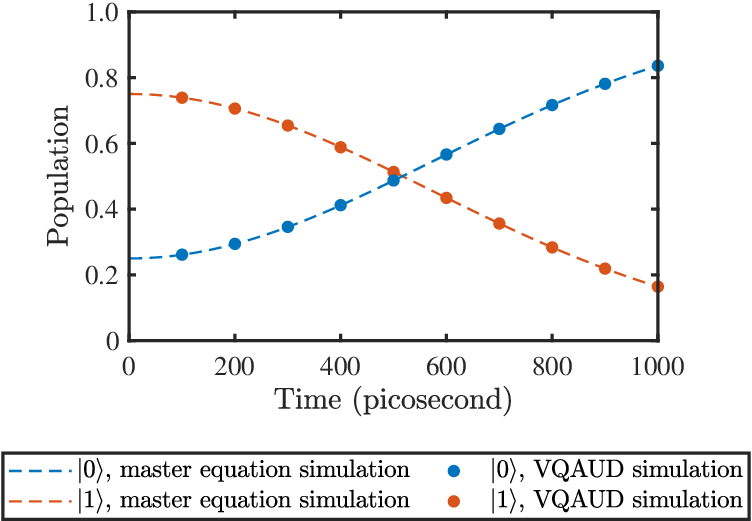}
\caption{Ideal VQAUD-based numerical simulation for the populations of the ground and excited states in a two-level open model ($\gamma=1.52\times10^9$~$\rm{s}^{-2}$) with Kraus method.}\label{Kraus}
\end{figure*}

In the main text, we have discussed in detail the application of VQAUD in implementing the vectorized Lindblad master equation for simulating open quantum systems. However, because of the extensive and complex nature of quantum open systems, alternative mathematical descriptions are sometimes required. Our approach is equally applicable to the Kraus operators approach. The dynamics of open quantum systems can be described through Kraus operators:
\begin{equation}
    \rho(t) = \sum_{k} M_k \rho M_k^\dagger.
\end{equation}
where $\{M_k\}$ satisfies the completeness condition $\sum_k M_k^\dagger M_k = I$. An initial $\rho$ is expressed as a sum of different pure quantum states weighted by the 
associated probabilities:

\begin{equation}
        \rho = \sum_i P_i |\phi_i\rangle\langle\phi_i|.
\end{equation}

This implies that by setting $M=M_k$ ($\alpha=1$), we can employ the VQAUD method to find an approximate unitary dilation $U$ of $M_k$, thereby allowing the simulation of the following equation in quantum circuits.

\begin{equation}
        U(\boldsymbol{\theta}^*)(|\phi_i\rangle \oplus |0\rangle) = (M_k|\phi_i\rangle) \oplus |\star\rangle.
\end{equation}
Each $M_k|\phi_i\rangle$ term corresponds to an independent quantum circuit. After obtaining the measurement outcome $M_k|\phi_i\rangle$ from each circuit, the complete $\rho(t)$ can be reconstructed through classical post-processing: $\rho(t)=\sum_{k}\sum_{i} P_iM_k |\phi_i\rangle\langle \phi_i| M_k^\dagger$.


We now concretely demonstrate the universal applicability of our approach for the Kraus operator method in the context of an open two-level model. In the operator sum representation:
\begin{equation}
\rho(t) = M_{0}\rho M_{0}^{\dagger} + M_{1}\rho M_{1}^{\dagger},
\end{equation}
where $M_0=\begin{pmatrix}
1 & 0 \\
0 & \sqrt{e^{-\gamma t^2}}
\end{pmatrix}$, and $M_1=\begin{pmatrix}
0 & \sqrt{1 - e^{-\gamma t^2}} \\
0 & 0
\end{pmatrix}$. 

Consider now an arbitrary initial $\rho=\frac{1}{4}\begin{pmatrix}
1 & 1 \\
1 & 3
\end{pmatrix}$ for an example that can be decomposed into $\rho=(|1\rangle \langle 1|+|+\rangle \langle +|)/2$, then the simulation result for the population of each basis is shown in Fig.~\ref{Kraus}.

\clearpage   
\bibliographystyle{apsrev4-2}

 \bibliography{SM.bbl}